\documentclass[aps,prl,showpacs,floatfix,twocolumn,longbibliography]{revtex4-2}
\usepackage{mathrsfs}
\usepackage[figuresright]{rotating}
\usepackage{amsmath}
\usepackage{amssymb}
\usepackage{graphicx}
\usepackage{color}
\usepackage{dcolumn}
\usepackage{bm}
\usepackage[breaklinks=true,colorlinks=true,linkcolor=blue,urlcolor=blue,citecolor=blue]{hyperref}
\usepackage{ragged2e}
\UseRawInputEncoding   
\usepackage{booktabs}
\usepackage{longtable}
\usepackage{multirow}
\usepackage{setspace}
\usepackage{soul}
\usepackage{xcolor}

\makeatletter

\newcommand{\Rmnum}[1]{\expandafter\@slowromancap\romannumeral #1@}
\newcommand{\supplementtableofcontents}{%
  \par
  \vspace{1em}
  \begin{center}
    \bfseries CONTENTS
  \end{center}
  \vspace{-0.5em}
  \@starttoc{smtoc}%
  \par
}
\newcommand{\SMsection}[1]{%
  \section{#1}%
  \addcontentsline{smtoc}{section}{%
    \protect\numberline{\thesection}#1%
  }%
}
\makeatother

\begin{document}

\title{Tilted $p$-wave magnet candidate CeNiAsO}

\author{Zhuo Wang$^{1}$}
\thanks{These authors contributed equally to this work.}
\author{Zheng Liu$^{2}$}
\thanks{These authors contributed equally to this work.}
\author{Shuo Zou$^{1}$}
\thanks{These authors contributed equally to this work.}
\author{Hua-Xun Li$^{3}$}
\author{Jin-Xin Hu$^{4}$}
\author{Zhuolun Qiu$^{1}$}
\author{Ze Wang$^{5}$}
\author{Jiamin Gong$^{6}$}
\author{Lucheng Wei$^{1}$}
\author{Kangjian Luo$^{1}$}
\author{Hai Zeng$^{1}$}
\author{Meng Zhang$^{1}$}
\author{Chao Dong$^{1}$}
\author{Chuanyin Xi$^{5}$}
\author{Junfeng Wang$^{1}$}
\author{Jiakun Fang$^{7}$}
\author{Xiaotao Han$^{1,7}$}
\author{Guang-Han Cao$^{3}$} \email[]{ghcao@zju.edu.cn}
\author{Liang Li$^{1,7}$} \email[]{liangli44@hust.edu.cn}
\author{Yongkang Luo$^{1,7}$} \email[]{mpzslyk@gmail.com}
\address{$^1$Wuhan National High Magnetic Field Center and School of Physics, Huazhong University of Science and Technology, Wuhan 430074, China;}
\address{$^2$Research Laboratory for Quantum Materials, Department of Physics and Materials, The Hong Kong Polytechnic University, Kowloon, Hong Kong, China;}
\address{$^3$School of Physics, Zhejiang University, Hangzhou, 310058, China;}
\address{$^4$Department of Physics, Hong Kong University of Science and Technology, Clear Water Bay, 999077 Hong Kong, China;}
\address{$^5$Anhui Key Laboratory of Low-Energy Quantum Materials and Devices, High Magnetic Field Laboratory, Hefei Institutes of Physical Science,
Chinese Academy of Sciences, Hefei 230031, China;}
\address{$^6$Aerospace Information Technology University, Jinan, Shandong 250200, P. R. China;}
\address{$^7$State Key Laboratory of Advanced Electromagnetic Technology, Huazhong University of Science and Technology, Wuhan 430074, China.}

\date{\today}

\begin{abstract}

The unexpectedly small ordered moments of CeNiAsO, a candidate for correlated $p$-wave magnet, have posed a serious challenge to the precise determination of its magnetic structure, hindering the understanding of its fundamental properties. By leveraging the high sensitivity to local internal fields, our $^{75}$As nuclear quadrupole / magnetic resonance experiments reveal a commensurate antiferromagnetic order with a small out-of-plane moment $m_z\approx0.05$ $\mu_{\mathrm{B}}$. This tilted magnetic configuration not only rotates the spin polarization axis away from the crystallographic $\mathbf{c}$-axis, but also enhances the non-relativistic spin splitting. We refer to this rare paradigm as a \textit{tilted $p$-wave magnet}.

\end{abstract}

\maketitle


Recently, altermagnetism has attracted widespread attention as a novel magnetic classification beyond conventional ferromagnetism and antiferromagnetism. Remarkably, altermagnetic materials exhibit spin-split electronic bands without showing net moment or stray field in real space \cite{Igor2022prx,Jungwirth2025Newton,LiborPhysRevX.12.031042,Libor2022PRX,Olena2024RuO2,RuO2_2022_prl,RuO2_prl2023,Zhou_2024prl_RuO2,Smejkal_2022,Jungwirth_Nature2026,MTe2024Nature}, holding great promise for spintronic applications. Typical examples include MnTe \cite{MTe2024Nature,Hariki_MnTe2024PRL,Yang_PRL2026}, CrSb \cite{Yang2025CrSb,Reimers2024CrSb,Santhosh2025CrSb,Zeng_AdvSci2024}, and RuO$_2$ \cite{LiborPhysRevX.12.031042, Olena2024RuO2, Libor2022PRX,RuO2_2022_prl,RuO2_prl2023,Zhou_2024prl_RuO2} that possess $d$- or $g$-wave symmetry. In contrast to these even-parity magnetisms in which the moments are collinearly ordered, an alternate branch -- odd-parity, $p$-wave magnetism -- often occurs in non-collinear orders, preserving the combined time-reversal and translation operation ($\mathcal{T\tau}$) symmetry \cite{P_wave_magnets_2023}. Such anisotropic nodal spin splitting potentially gives rise to even richer properties and application prospects, e.g. Majorana fermions via proximity coupling to superconductors \cite{EzawaM-PRB2024, MaedaK-JPSJ2024,Sukhachov_PRB2025,Sun_PRB2025,Fukaya_JPCM2025}, as well as the highly efficient spin-charge conversion in the absence of spin-orbit coupling -- namely, the non-relativistic Edelstein effect \cite{Chakraborty2025_NC,Brekke_prl2024,Gu_PRL2025}.

Among the handful of proposed candidates for $p$-wave magnet \cite{NiI2_2025Nature,Gd3Ru4Al12_nature,P_wave_magnets_2023,ZhaoPRB2025}, the heavy-fermion compound CeNiAsO is special in that the strong electronic correlation effect arising from the fluctuating Ce-$4f$ electrons likely endow it with higher tunability and better functionality. CeNiAsO crystallizes in the tetragonal ZrCuSiAs-type structure with the space group $P4/nmm$ (No. 129), iso-structural to the well-known 1111 iron-based superconductors \cite{Hosono-LaOFFeAs}. The cerium moments in this compound undergo two sequential antiferromagnetic (AFM) transitions at $T_\text{N1}\sim9$ K and $T_\text{N2}\sim7$ K, respectively \cite{YongkangLuo2011}. Unconventional heavy-fermion quantum criticality and strange-metal-like behavior were attained by either pressure tuning or As-for-P doping \cite{luo2014}. Neutron scattering and muon spin rotation studies suggested that its ground-state magnetic structure for below $T_\text{N2}$ is a commensurate AFM (CAFM) order with propagation vector $\mathbf{q} = (0.5,~0,~0)$ and $|\pm 1/2\rangle$ Kramers doublet character \cite{ShanWu2019}. This proposed magnetic structure is characterized by a \textit{coplanar}, non-collinear arrangement of moments, preserving both $\mathcal{T\tau}$ and ${C}_{2z}^s\mathcal{T}$ symmetries, with the spin polarization oriented perpendicular to the moment plane ($\mathbf{S}_z \parallel \mathbf{c}$). These features prompted the suggestion of candidate $p$-wave magnet \cite{P_wave_magnets_2023, Chakraborty2025_NC, ZhaoPRB2025, LuoX-OddParityMag}.

However, many aspects of this compound remain puzzling. First of all, the ordered moment ($\sim0.37~\mu_\text{B}$/Ce \cite{ShanWu2019}) is far less than the expected value for $|\pm 1/2\rangle$ doublet and than that in the structural analog CeFeAsO ($0.83 ~\mu_\text{B}$/Ce) \cite{ZhaoJ-CeFeAsO_F2008}. Second, this magnetic order is highly resistant to external field -- the $\mathbf{c}$-axis magnetization reaches only $0.17~\mu_\text{B}$/Ce up to 58~T (see Fig.~S1 in \textbf{Supplemental Material} (\textbf{SM}) \cite{SM}). Although Kondo hybridization might plausibly explain the reduction of Ce moments, this is ruled out by the high magnetic entropy gain ($>0.7R\ln2$) at $T_\text{N1}$ \cite{YongkangLuo2011} and the recent angular resolved photoemission spectroscopy (ARPES) experiments \cite{ZhangJ-CeNiAsO_ARPES,zhangX-CeNiAsO_ARPES}, which all confirm that Ce-$4f$ is highly localized. Last but not least, the mysteriously small magnetic moment also imposes exceptional difficulty to definitively resolve $m_z$, the out-of-plane component of the Ce moments \cite{ShanWu2019,LuFangjunCeNiAsO}. Whether the moments are strictly coplanar, and how the $z$-canted components are arranged, directly dictate the preservation or breakdown of the ${C}_{2z}^s\mathcal{T}$ symmetry. This epitomizes a broader, unresolved issue in the field of altermagnetism: how does this canting universally impact the stability of symmetry-enforced spin-splitting? Accurately resolving the magnetic structure of CeNiAsO, therefore, provides an essential benchmark for disentangling these competing effects and lays a robust foundation for understanding the underlying physical properties \cite{Zhou_2025arXiv_CeNiAsO, ZhangF-CeNiAsO_Odd, ZhangJ-CeNiAsO_ARPES,zhangX-CeNiAsO_ARPES}.

\begin{figure}[!ht]
\vspace*{-0pt}
\hspace*{-0pt}
\includegraphics[width=8.5cm]{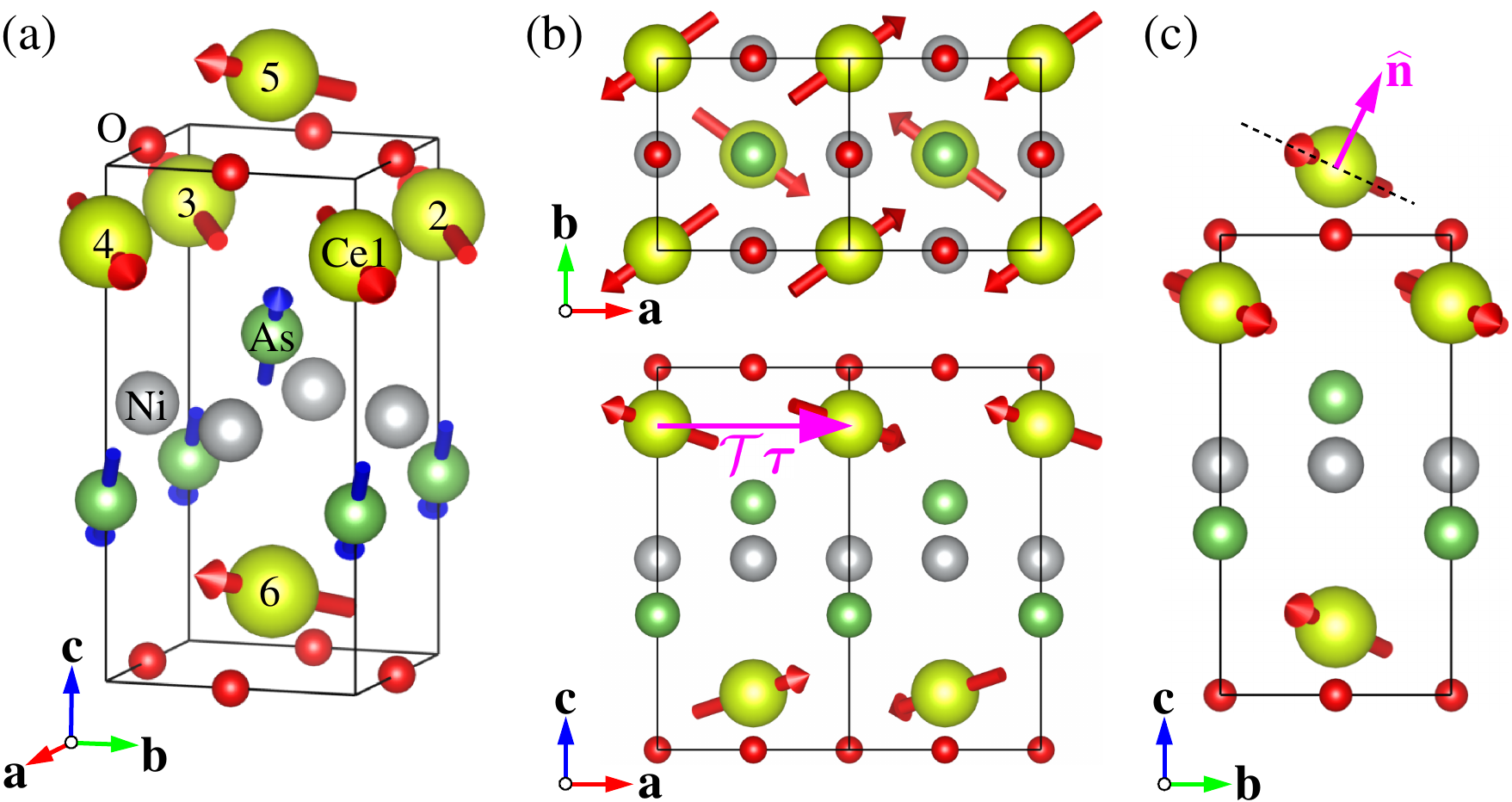}
\vspace*{-15pt}
\caption{(a) Crystalline / magnetic structure of CeNiAsO, and local environment of As. The red arrows denote the proposed commensurate AFM structure of Ce ions with $\mathbf{q}=(0.5, 0, 0)$ for $T<T_\text{N2}$, while the blue arrows indicate the internal field at the As sites. (b) Top-view and side-view of the magnetic structure; the canted component $m_z\sim 0.05$ $\mu_\text{B}$. $\mathcal{T}\tau$ symmetry is preserved in this magnetic structure. (c) Tilted spin polarization axis $\hat{\mathbf{n}}\approx (0, ~0.231, ~0.973)$. }
\label{Fig1}
\end{figure}

Here, by virtue of the high-sensitivity to local internal field, we exploit $^{75}$As nuclear quadrupole resonance (NQR) and nuclear magnetic resonance (NMR) to reinvestigation into the magnetic order of CeNiAsO. Our analysis reveals that the ground state of CeNiAsO is a tilted coplanar CAFM order with a small $\mathbf{c}$-canting moment $m_z\approx0.05$ $\mu_{\mathrm{B}}$. This magnetic canting establishes a new spin plane and reconstructs the spin-group symmetry, resulting in a reorientation of the spin polarization axis $\hat{\mathbf{n}}$ deviating from the crystallographic $\mathbf{c}$ axis (Fig.~\ref{Fig1}). The anomalous Hall effect (AHE) is forbidden at zero field by the preserved $\mathcal{T}\tau$ symmetry, but emerges under a high magnetic field when this symmetry is broken, revealing the intimate connection between magnetic-symmetry reconstruction and the $p$-wave spin-splitting behavior, which is further supported by first-principles calculations.

$^{75}$As (nuclear spin $^{75}I$ = 3/2, gyromagnetic ratio $^{75}\gamma_n$ = 7.2919 MHz/T, and quadrupole moment $^{75}Q=0.29$ barns) NQR and NMR (external field $\mu_0H=10$ T) measurements were conducted on high-quality CeNiAsO samples. High-field Hall effect and magnetization measurements were made at Hefei (steady) and Wuhan (pulsed) high magnetic field laboratories, respectively. $\text{DFT}+U$ calculations \cite{Anisimov-DFT+U_PRB1991} with constrained moment directions \cite{Ma-PRB2015} were performed using Vienna \textit{ab initio} simulation package (VASP) \cite{Kresse-VASP_PRB1996} with Perdew-Burke-Ernzerhof type generalized gradient approximation \cite{Perdew-PBE_PRB1992} and projected augmented-wave \cite{Blochl-PAW_PRB1994} pseudo-potentials. The anomalous Hall conductivity was then evaluated via Wannier functions \cite{Mostofi-Wannier90_CPC2008}. See \textbf{SM} for details \cite{SM}.

\begin{figure*}[!htp]
\vspace{0pt}
\hspace{-05pt}
\includegraphics[width=18.0cm]{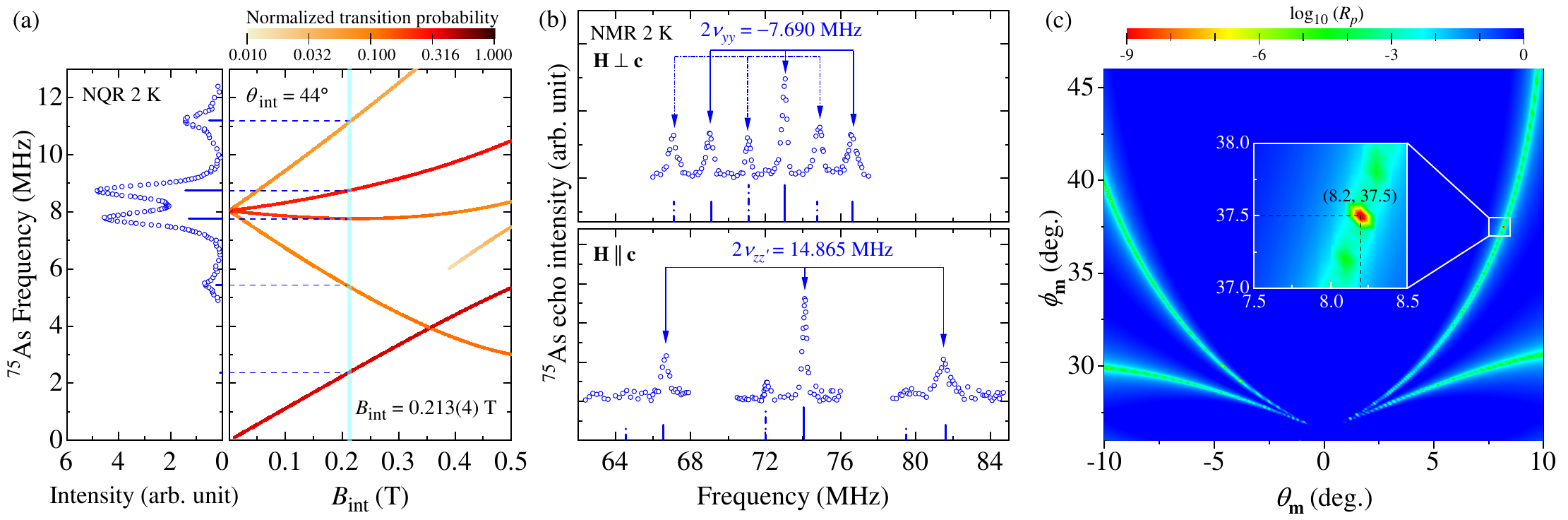}
\vspace{-18pt}
\caption{(a) Comparison of the experimental (left) and calculated (right) $^{75}$As NQR spectra determines the angle between $\mathbf{B}_\text{int}$ and the EFG principle axis, $\theta_\text{int}\approx44~^\circ$, and $B_\text{int}\approx0.213$ T. The colormap represents the normalized transition probability. (b) $^{75}$As NMR spectra of single crystalline CeNiAsO for $\mathbf{H}\perp\mathbf{c}$ (top) and $\mathbf{H}\parallel\mathbf{c}$ (bottom). Note that $\nu_{zz'}$ differs slightly from $\nu_{zz}$ (or $\nu_Q$) due to the small misalignment of $\mathbf{c}$ axis with respect to $\mathbf{H}$. The vertical lines in these panels signify the simulated NMR frequencies. 
(c) The contour plot of $\log_{10}R_p$ gives the optimal parameters for the refinement, $\theta_\mathbf{m}=8.2~^\circ$, and  $\phi_\mathbf{m}=37.5~^\circ$. }
\label{Fig2}
\end{figure*}

The raw data of $^{75}$As NQR spectra of CeNiAsO at low temperature are displayed in Fig.~S2(a) \cite{SM}. For $^{75}I$ = 3/2, only one resonance peak is expected in the NQR spectra in the paramagnetic regime, which is indeed the case at 10 K. $\nu_Q$ = 8.007 MHz is obtained by Gaussian fitting, close to but slightly larger than that from NMR experiments at 10 T (Fig.~S3). Below $T_\text{N1}$, for instance 8 K and 7.5 K, the peak broadens significantly due to the continuous distribution of the internal field in the incommensurate AFM (ICAFM) order. Upon further cooling to below $T_\text{N2}$, the NQR peak clearly splits into two peaks [Fig.~\ref{Fig2}(a)], manifesting the development of CAFM order \cite{ShanWu2019, LuFangjunCeNiAsO}.

Such an evolution of magnetic order in CeNiAsO is also evidenced by $^{75}$As NMR experiments. In Fig.~S2(b-e) \cite{SM}, we reproduce our previous aligned-powder NMR spectra in gray \cite{LuFangjunCeNiAsO}, and show the recent single crystalline NMR results in red (10 K) and blue (2 K). On the whole, the single crystal results well resemble those from aligned powders. At 10 K in the paramagnetic phase, three distinct peaks are discernible for both $\mathbf{H}\parallel\mathbf{c}$ and $\mathbf{H}\perp\mathbf{c}$, arising from the central ($-\frac{1}{2}\leftrightarrow\frac{1}{2}$) and satellite ($\pm\frac{1}{2}\leftrightarrow\pm\frac{3}{2}$) transitions of the $^{75}I=3/2$ nucleus. Further cooled down to below $T_\text{N2}$, each peak splits [seeing also Fig.~\ref{Fig2}(b)], which is better seen in the single crystal data, characteristic of CAFM order.

We now analyze these NQR and NMR results in depth aiming to obtain the magnetic structure of the CAFM phase. For more details, we refer the readers to $\textbf{SM}$ \cite{SM}. First of all, it is necessary to figure out the internal field ($\mathbf{B}_\text{int}$) with respect to the electric field gradient (EFG), and this can be obtained by a combination of numerical simulation and fitting to the experimental spectra. To account for both NQR and NMR results, we write the total Hamiltonian as
\begin{equation}    \mathcal{H}_\text{tot}=\mathcal{H}_\text{Z}+\mathcal{H}_\text{int}+\mathcal{H}_\text{Q},
    \label{Eq.Htot}
\end{equation}
where $\mathcal{H}_\text{Z}$ is the Zeeman term caused by the external field; $\mathcal{H}_\text{int}$ is the (Zeeman) term induced by internal field; and $\mathcal{H}_\text{Q}$ is the nuclear quadrupole interaction with EFG \cite{Slichter}, taking the form
\begin{equation}
\mathcal{H}_Q=\frac{eQV_{zz}}{4I(2I-1)}[3\hat{I_z}^2-\mathbf{\hat{I}}^2+\eta(\hat{I_x}^2-\hat{I_y}^2)],
\label{Eq.HQ}
\end{equation}
in which $\mathbf{\hat{I}}=(\hat{I}_x,~\hat{I}_y,~\hat{I}_z)$ is nuclear spin operator, and $\eta\equiv(V_{xx}-V_{yy})/V_{zz}$ is the asymmetry parameter with $V_{xx}$, $V_{yy}$ and $V_{zz}$ being the components of the EFG tensor. First, to simulate the zero-field NQR results, we ignore $\mathcal{H}_\text{Z}$ in Eq.~(\ref{Eq.Htot}), and calculate the resonance frequencies as functions of $B_\text{int}$ with varying polar angle $\theta_\text{int}$. The representative results are provided in Fig.~S5 \cite{SM}. We find the results that best fit our NQR spectra are $B_\text{int}=0.213$ T with $\theta_\text{int}=44~^\circ$, as shown in Fig.~\ref{Fig2}(a). Second, the azimuthal angle $\phi_\text{int}$ of the $\mathbf{B}_\text{int}$ can also be determined by calculating the central peak splitting  ($\Delta f$) of the $^{75}$As NMR spectra, as shown in Figs.~S4 and S6 \cite{SM}. We find $\phi_\text{int}=26.5~^\circ$ uniquely fits the experimental satisfactorily. Hence, the internal field at the As site can be fully determined, as is shown by the blue arrows in Fig.~\ref{Fig1}(a).

In Ce-based heavy-fermion compounds, the internal field is mainly governed by the transferred hyperfine interaction with Ce-$4f$ spins that relies on orbital overlap \cite{Curro_2006CeMIn5, Curro2004Twofluid}. Note that Ni is non-magnetic in CeNiAsO and thus has little contribution to spin susceptibility \cite{GXu2008,FRonning-Ni2X2,YongkangLuo2011}. In this context, each As has four nearest-neighbor Ce spins [labeled as Ce1-4 in Fig.~\ref{Fig1}(a)]. In addition, the two next-nearest-neighbor Ce5 and Ce6 -- which sit right above and below As -- should also be taken into account due to their considerable hybridization with As-$4p_z$ orbitals \cite{LuFangjunCeNiAsO}. Therefore, we have
\begin{equation}
\mathbf{B}_\text{int}=\sum_{i=1}^6\mathbb{A}_i \cdot \mathbf{m}_i,
\label{EqBint}
\end{equation}
where $\mathbb{A}_i$ is the $i$-th hyperfine tensor, and $\mathbf{m}_i$ is the ordered magnetic moment of the $i$-th Ce ion. The detailed forms of $\mathbb{A}_i$ tensors are described in \textbf{SM} \cite{SM}. To obtain the magnetic structure, we combine the internal field information obtained above and use the magnetic moment $m\approx0.37$ $\mu_\text{B}$ from previous neutron scattering studies \cite{ShanWu2019}. For different orientations of ordered moments characterized by a combination of polar angle ($\theta_\mathbf{m}$) and azimuthal angle ($\phi_\mathbf{m}$), hyperfine coupling constant $A_{\text{hf},ab}^\text{cal}$ can be computed and compared with the experimental $A_{\text{hf},ab}^\text{exp}=-0.42$ T/$\mu_\text{B}$ \cite{LuFangjunCeNiAsO}. The goodness of the fitting is evaluated by
\begin{equation}
R_p(\theta_\mathbf{m},~\phi_\mathbf{m})={(A_{\text{hf},ab}^\text{cal}-A_{\text{hf},ab}^\text{exp})^2}.
\label{EqRp}
\end{equation}
The optimal parameters $\theta_\mathbf{m}=8.2~^\circ$ and $\phi_\mathbf{m}=37.5~^\circ$ are obtained where $\log_{10}R_{p}$ minimizes, as is shown in Fig.~\ref{Fig2}(c). Note that the derived $\phi_\mathbf{m}$ is very close to $36~^\circ$, the value reported by the previous neutron scattering study \cite{ShanWu2019}. The little $\theta_\mathbf{m}$ yields a small $m_z\approx0.05~\mu_\text{B}$, which is also within the uncertainty of the neutron scattering experiment ($\sim 0.06~\mu_\text{B}$) \cite{ShanWu2019}. The as-proposed canted CAFM order is depicted in Fig.~\ref{Fig1}, and a comparison of measured and simulated NMR spectra is provided in Fig.~\ref{Fig2}(b). The sign of $m_z$ alternates along the $\mathbf{a}$ axis in this magnetic structure; we hereafter denote it as CAFM$_\text{z}$ for brevity.

\begin{figure}[!htp]
\vspace{0pt}
\hspace{-3pt}
\includegraphics[width=8.7cm]{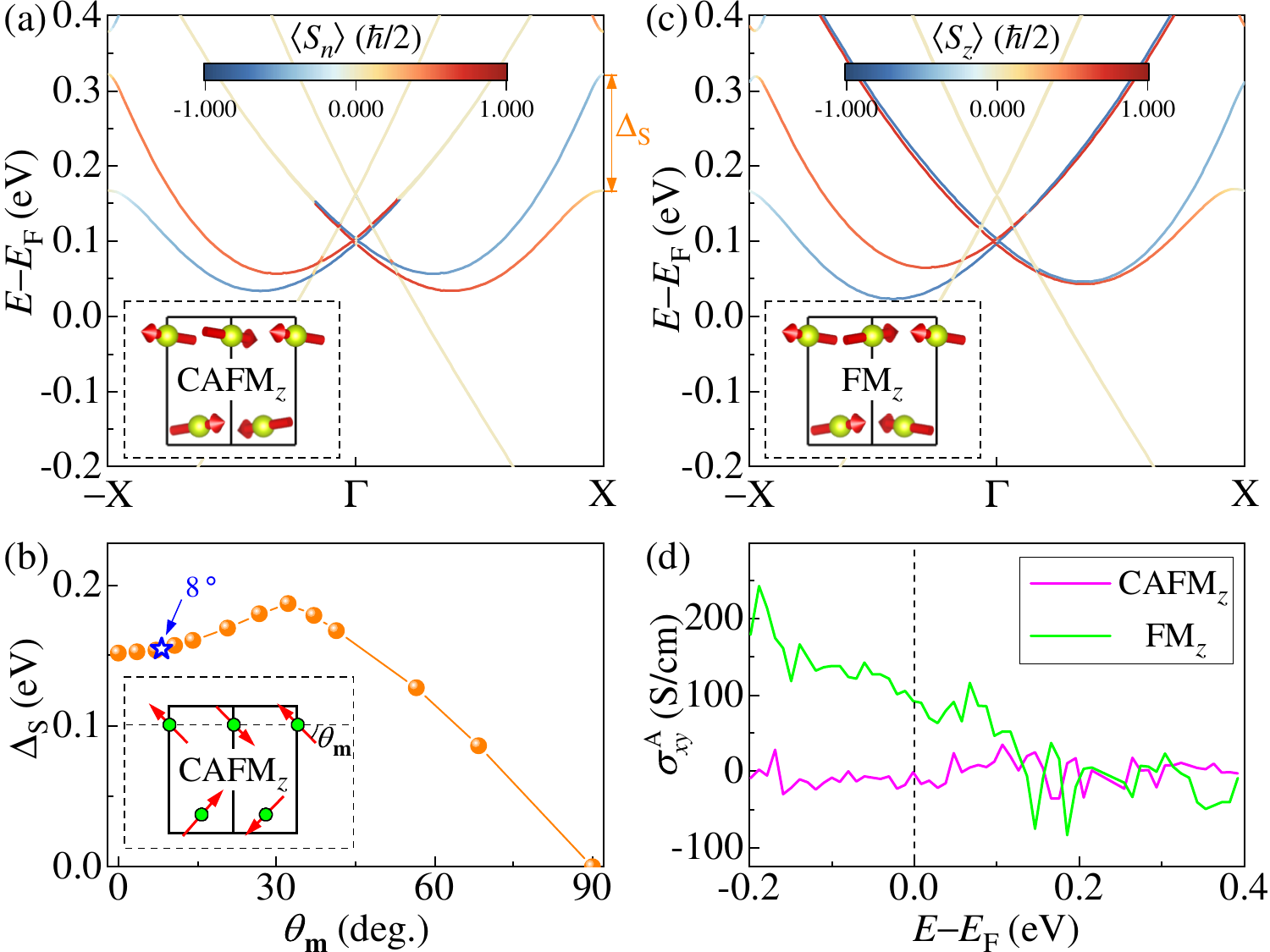}
\vspace{-15pt}
\caption{(a) Spin dependent band structure of CeNiAsO for CAFM$_z$, $\theta_\mathbf{m}=8~^\circ$. (b) Spin band splitting $\Delta_\text{S}$ as a function of $\theta_\mathbf{m}$ in CAFM$_z$. (c) Spin dependent band structure for canted FM$_z$, $\theta_\mathbf{m}=8~^\circ$; this configuration is to simulate the oriented magnetic state under high field $\mathbf{H}\parallel\mathbf{c}$. (d) Calculated anomalous Hall conductivity $\sigma_{xy}^\text{A}$ for CAFM$_z$ and FM$_z$ configurations with $\theta_\mathbf{m}=8~^\circ$.}
\label{Fig3}
\end{figure}

Due to the emergence of a finite $m_z$ component, the $C_{2z}^s\mathcal{T}$ symmetry is violated, and the spin polarization is no longer constrained along the crystallographic $\mathbf{c}$-axis. Instead, the canted magnetic moments define a new spin plane with a normal vector $\hat{\mathbf{n}}\approx(0,~0.231,~0.973)$ \cite{SM}. Consequently, the spin-group symmetry is reconstructed into a combined operation of $\mathcal{T}$ and a twofold spin rotation about $\hat{\mathbf{n}}$, namely $C_{2n}^s\mathcal{T}$. Under this symmetry, together with the preserved $\mathcal{T}\tau$ symmetry, the spin polarization axis of the electronic bands is rotated from the crystallographic $\mathbf{c}$-axis to the magnetic axis $\hat{\mathbf{n}}$, resulting in a reorientation of the altermagnetic spin polarization. To see this clearly, we performed first-principles calculations using the $\text{DFT}+U$ method. Relative large $U=5$ eV and $J=2$ eV were employed to account for the localized Ce-$4f$ electron here \cite{ZhangJ-CeNiAsO_ARPES, zhangX-CeNiAsO_ARPES}. Figure~\ref{Fig3}(a) shows the band structure along the $\Gamma$-$\text{X}$ path. Despite the tilted CAFM$_z$ structure, the non-relativistic odd-parity spin-split bands remain robustly preserved. 
Remarkably, by calculating the spin band splitting $\Delta_\text{S}$ as a function of the canting angle $\theta_\mathbf{m}$, we uncover a non-monotonic dependence as shown in Fig.~\ref{Fig3}(b), \textit{i.e.}, $\Delta_\text{S}$ initially increases with $\theta_\mathbf{m}$, reaching a maximum near 32 $^\circ$ where its value is enhanced by $23~\%$ as compared with that in the non-canted case, and then decreases continuously and diminishes at $\theta_\text{m}=90~^\circ$. Our results demonstrate that magnetic tilting provides a new route to actively reorient the spin polarization axis and to further promote the non-relativistic Edelstein effect in altermagnetic materials, beyond the conventional picture that the spin texture is constrained by the crystallographic symmetry.

Another hallmark for $p$-wave magnet is the absence of zero-field AHE, prohibited by the $\mathcal{T}\tau$ symmetry that enforces $\boldsymbol{\Omega}(\boldsymbol{k})=-\boldsymbol{\Omega}(-\boldsymbol{k})$,
leading to a vanishing Berry curvature ($\boldsymbol{\Omega}$) integral over the Brillouin zone. This is indeed the case in the CAFM$_z$ configuration of CeNiAsO, seeing Fig.~\ref{Fig3}(d). Nonetheless, the AHE can be expected to emerge when the magnetic moments are reoriented by an external field, provided that the $T\tau$ symmetry is broken. To see this, we also simulate a FM$_\text{z}$ configuration wherein the magnetic moments have a partially polarized component along $\mathbf{c}$ [Fig.~\ref{Fig3}(c)]; a significant $\sigma_{xy}^\text{A}$ is anticipated in this configuration, cf Fig.~\ref{Fig3}(d).

\begin{figure}[!htp]
\vspace{0pt}
\hspace{-0pt}
\includegraphics[width=8.8cm]{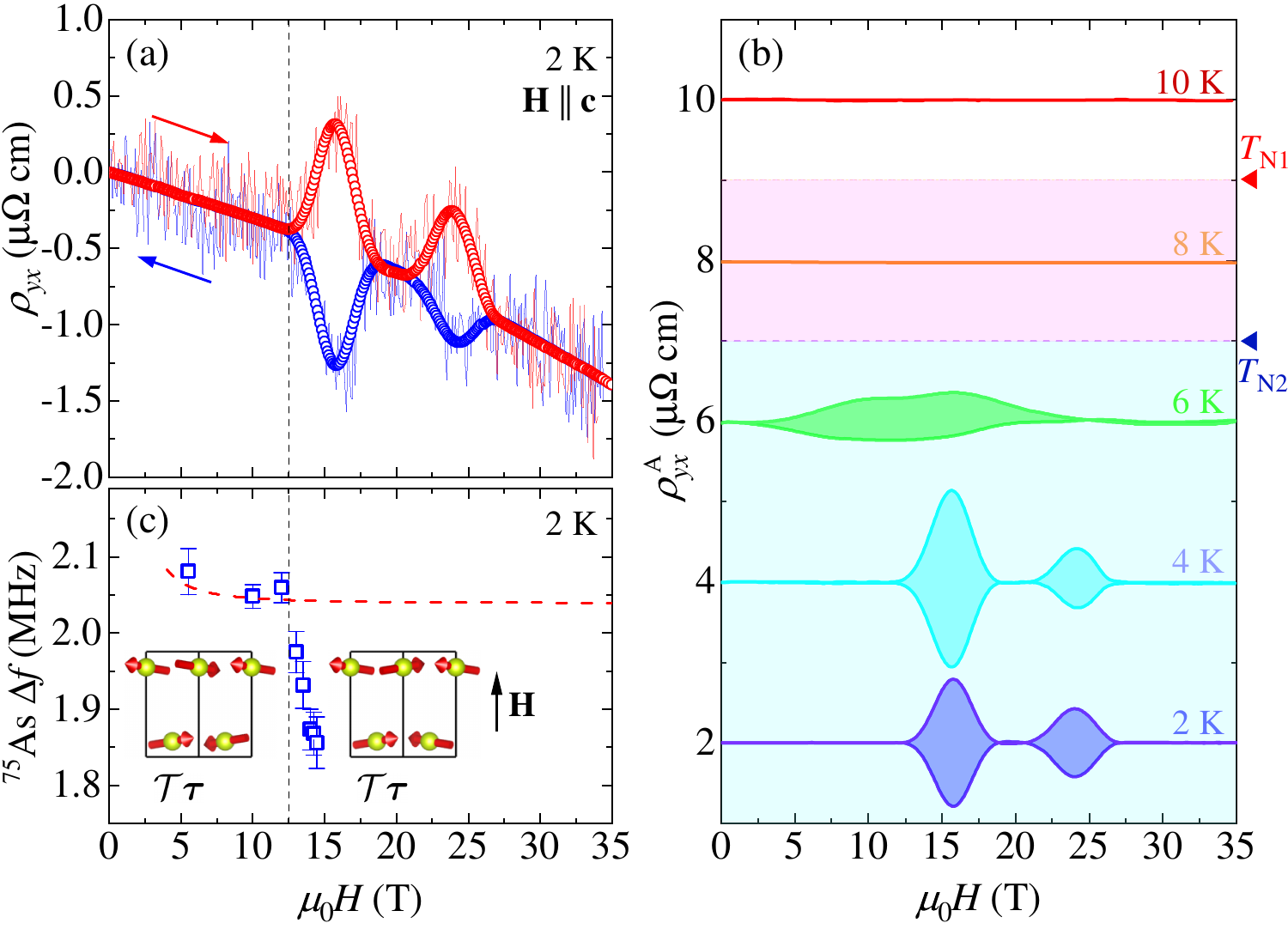}
\vspace{-15pt}
\caption{(a) Hall resistivity as a function of field at 2 K. The open symbols represent the smoothed curve, serving as guides to the eye. (b) AHE at various temperatures (raw data in Fig.~S8 \cite{SM}). The curves are vertically shifted for clarity. The colors of the background manifest distinct magnetic phases: white - paramagnetic; magenta - ICAFM \cite{ShanWu2019}; cyan - CAFM. (c) $^{75}$As NMR central peak splitting $\Delta f$ measured under varying field. The red dashed line is a theoretical prediction assuming that the Ce moments do not reorient with applied field. The insets depict the conjectured magnetic structures in the field regions below and above $\sim 13$ T, wherein the $\mathcal{T\tau}$ symmetry is preserved and violated, respectively.}
\label{Fig4}
\end{figure}

The experimental observation of this evolution, however, faces formidable challenges, primarily due to the good metallicity and the exceptionally hard magnetization of CeNiAsO (Fig.~S1 \cite{SM}). To overcome these difficulties, we performed high-precision high-field Hall effect measurements with $\mathbf{H}\parallel \mathbf{c}$, as shown in Fig.~\ref{Fig4} and Fig.~S8 \cite{SM}. At 2 K, the Hall resistivity $\rho_{yx}$ consists of two parts: an ordinary contribution $\rho_{yx}^\text{O}$, and a loop-structured AHE $\rho_{yx}^\text{A}$. The ordinary part can be well fitted to $\rho_{yx}^\text{O}(H)=R_\text{H} H + R_2 H^2$; the $H^2$ term is expected to become important in the high-field regime where the condition $\omega_\text{c}\tau_\text{tr}\ll 1$ no longer holds ($\omega_c$ is cyclotron frequency; $\tau_\text{tr}$ is scattering time) \cite{Nair-CeIrIn52008, HeX-Ce3TiSb5_Hall2025}. The obtained small Hall coefficient $R_\text{H}=2.9(1)\times 10^{-10}$ m$^3$/C reaffirms the good metallicity \cite{luo2014}. The AHE component is extracted and displayed in Fig.~\ref{Fig4}(b) for temperatures in different magnetic phases. Remarkably, such AHE appears only in the CAFM$_z$ phase, and disappears immediately when $T$ surpasses $T_\text{N2}$. Moreover, by checking the field dependent $^{75}$As NMR spectra [Fig.~S4(a)] \cite{SM}, one finds that the central peak splitting $\Delta f$ only slightly decreases for field below 12 T, but starts to drop rapidly when field exceeds 13 T, cf Fig.~\ref{Fig4}(c); this onset field agrees well with the field region where the first AHE loop is observed (black vertical dashed line). Such a substantial reduction of $\Delta f$ is a signature of the reorientation of Ce moments under high field [seeing the red dashed line in Fig.~\ref{Fig4}(c)]. We conjecture that once field is large enough to generate a net $m_z$ (viz FM$_z$ state), the $\mathcal{T\tau}$ symmetry is broken, and this can lead to the appearance of AHE. Due to the small magnetic moment and the extremely-hard magnetization process [Fig.~S1(c)] \cite{SM}, such a field-induced reorientation is expected to be slow or quasi-continuous, for which magnetization measurements under pulsed high field may not be capable of resolving this feature. We should also admit that the underlying mechanism for the two separate AHE loops at low temperature remains unclear to us. To further clarify this issue, additional NMR experiments with much higher field will be necessary. We leave this open question for future investigation.


In conclusion, we successfully revised the magnetic structure of CeNiAsO utilizing a combination of $^{75}$As NQR and NMR measurements. The results reveal a tilted-coplanar commensurate AFM order, featuring a small out-of-plane moment $m_z\approx 0.05~\mu_\text{B}$ that readily breaks the $C_{2z}^s\mathcal{T}$ symmetry while preserves the $\mathcal{T\tau}$ symmetry, thereby giving rise to potential tilted $p$-wave magnetism, which is further supported by $\text{DFT}+U$ calculations and the avoided AHE at zero field. Our work prompts a renewed examination of the peculiar $p$-wave magnetism and the underlying physical properties of this heavy-fermion altermagnet. In addition, the method we established in this work should also be applicable to determining the magnetic structure in other small-ordered-moment systems, particularly under the extreme conditions (e.g. high pressure or strong magnetic field) where conventional neutron techniques may prove inadequate.


\textit{Acknowledgments}--The authors acknowledge Joe D. Thompson and Wu Xie for helpful discussions. This work is supported by National Key R\&D Program of China (2023YFA1609600, 2022YFA1602602, 2023YFA1406101), National Natural Science Foundation of China (U23A20580 and 52588101), and Beijing National Laboratory for Condensed Matter Physics (2024BNLCMPKF004).
We thank the HM of the Steady High Magnetic Field Facility, CAS (https://cstr.cn/31125.02.SHMFF.HM) for the assistance on the experiment. We also thank the Magnetic Property Station of Wuhan National High Magnetic Field Center (https://whmfc.hust.edu.cn/yhfw/cljs.htm).


\textit{Data availability}--All data that support the findings of this study are available from the corresponding authors upon request.


%

\newpage

\renewcommand{\thefigure}{S\arabic{figure}}
\renewcommand{\thetable}{S\arabic{table}}
\renewcommand{\theequation}{S\arabic{equation}}
\onecolumngrid
\newpage
\begin{center}
{\bf \large
{Supplemental Material:}\\
Tilted $p$-wave magnet candidate CeNiAsO
}
\end{center}

\setcounter{table}{0}
\setcounter{figure}{0}
\setcounter{equation}{0}
\setcounter{page}{1}
\small
\begin{center}
Zhuo Wang$^{1*}$, Zheng Liu$^{2*}$, Shuo Zou$^{1*}$, Hua-Xun Li$^{3}$, Jin-Xin Hu$^{4}$, Zhuolun Qiu$^{1}$, Ze Wang$^{5}$, Jiamin Gong$^{6}$, Lucheng Wei$^{1}$, Kangjian Luo$^{1}$, Hai Zeng$^{1}$, Meng Zhang$^{1}$, Chao Dong$^{1}$, Chuanyin Xi$^{3}$, Junfeng Wang$^{1}$, Jiakun Fang$^{7}$, Xiaotao Han$^{1,7}$, Guang-Han Cao$^{2\dag}$\email{ghcao@zju.edu.cn}, Liang Li$^{1,7\ddag}$\email{liangli44@hust.edu.cn}, and Yongkang Luo$^{1,7\S}$\email{mpzslyk@gmail.com}\\
$^1${\it Wuhan National High Magnetic Field Center and School of Physics, Huazhong University of Science and Technology, Wuhan 430074, China;}\\
$^2${\it Research Laboratory for Quantum Materials, Department of Physics and Materials, The Hong Kong Polytechnic University, Kowloon, Hong Kong, China;}\\
$^3${\it School of Physics, Zhejiang University, Hangzhou, 310058, China;}\\
$^4${\it Department of Physics, Hong Kong University of Science and Technology, Clear Water Bay, 999077 Hong Kong, China;}\\
$^5${\it Anhui Key Laboratory of Low-Energy Quantum Materials and Devices, High Magnetic Field Laboratory, Hefei Institutes of Physical Science, Chinese Academy of Sciences, Hefei 230031, China;} \\
$^6${\it Aerospace Information Technology University, Jinan, Shandong 250200, P. R. China;}\\
$^7${\it State Key Laboratory of Advanced Electromagnetic Technology, Huazhong University of Science and Technology, Wuhan 430074, China.}\\

\date{\today}
\end{center}
\normalsize
\vspace*{15pt}

In this \textbf{Supplemental Material} (\textbf{SM}), we provide additional results that further support the discussion and conclusion in the main text, including experimental and calculation methods, sample characterization, magnetization under pulsed high field, Hall effect, the raw $^{75}$As NQR and NMR data, and the details of internal field calculation. \\

\emph{}\\
\emph{}\\
\emph{}\\





\supplementtableofcontents

\newpage

\SMsection{SM \Rmnum{1}. M\lowercase{ethods}}

\begin{figure}[!htp]
\vspace{-0pt}
\hspace{-0pt}
\includegraphics[width=16cm]{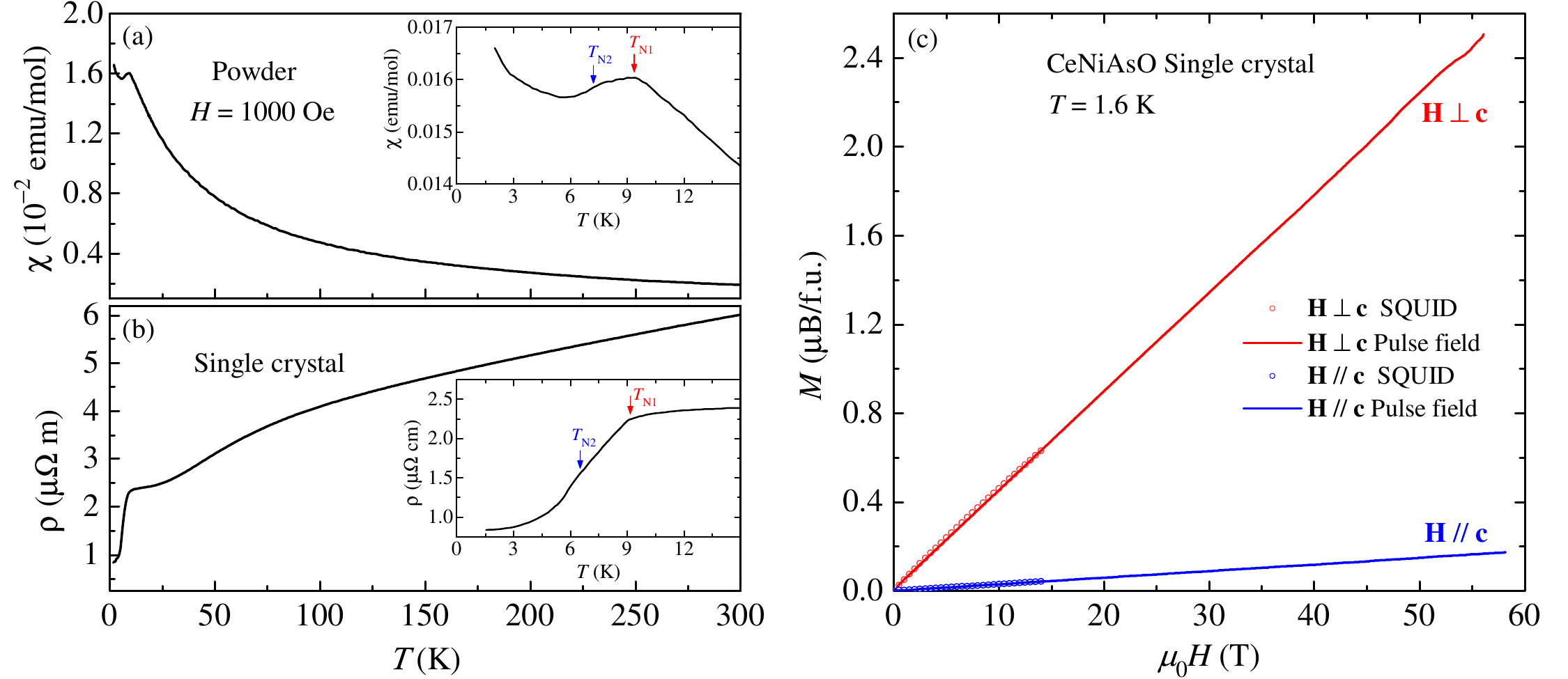}
\vspace{-0pt}
\caption{Sample characterization of CeNiAsO. (a) Temperature dependence of magnetic susceptibility, $\chi(T)$. (b) Electrical resistivity as a function of $T$, $\rho(T)$. The two AFM transitions are seen in both polycrystalline and single-crystalline samples. (c) Isothermal field dependent magnetization, $M(H)$, measured at 1.6 K. The magnetizations under pulsed field are calibrated by the low-field results measured with SQUID. For both $\mathbf{H}\parallel\mathbf{c}$ and $\mathbf{H}\perp\mathbf{c}$, $M(H)$ curves are linear with field up to 58 T. In particular, $M_c$ reaches only 0.17 $\mu_\text{B}$ at 58 T.}
\label{FigS1}
\end{figure}

High-purity polycrystalline CeNiAsO was grown by solid-state reaction \cite{YongkangLuo2011, luo2014}. CeNiAsO single crystals were grown by the NaAs-flux method as described elsewhere \cite{Zhou_2025arXiv_CeNiAsO}. The quality of these samples were characterized by magnetic susceptibility and resistivity measurements (Fig.~\ref{FigS1}). 
$^{75}$As (nuclear spin $^{75}I$ = 3/2, gyromagnetic ratio $^{75}\gamma_n$ = 7.2919 MHz/T, and quadrupole moment $^{75}Q=0.29$ barns) NQR and NMR spectra were acquired by a stepped frequency-sweep method using a Redstone spectrometer (Tecmag) and a high-homogeneity superconducting magnet (Oxford Instruments). Approximately 500 mg of powder sample was used for the zero-field NQR measurements. For NMR experiments, 19 pieces of CeNiAsO single crystals with a total mass $\sim$2 mg were accumulated and properly aligned in the sample coil; the measurements were made under $\mu_0H\approx10$ T, the precise value of which was determined by the in-situ $^{63}$Cu shift of the sample coil. 
High-field Hall effect and magnetization measurements were conducted at Hefei (steady) and Wuhan (pulsed) high magnetic field laboratories, respectively. High-precision Hall signal was measured by a low-noise lock-in amplifier SR865A equipped with an SR554A pre-amplifier.

Our first-principles calculations were performed by using the projected augmented-wave method \cite{Blochl-PAW_PRB1994} as implemented in the Vienna \textit{ab initio} simulation package (VASP) \cite{Kresse-VASP_PRB1996}. The generalized gradient approximation of the Perdew-Burke-Ernzerhof type was used to describe the exchange-correlation interaction \cite{Perdew-PBE_PRB1992}. All the atoms were allowed to relax until the Hellmann-Feynman force on each atom is smaller than 0.02 eV/\AA. The $\Gamma$-centered Monkhorst-Pack grid of $4\times8\times4$ was used during the calculation. The plane-wave energy cutoff was set to 460 eV. The $\text{DFT}+U$ method \cite{Anisimov-DFT+U_PRB1991} with $U=5.0$ eV and $J=2.0$ eV was used to account for the on-site Coulomb interaction of the Ce-$4f$ electrons. The directions of the magnetic moments were constrained by adding a penalty functional to the total energy \cite{Ma-PRB2015}. The Wigner-Seitz radii were chosen as 1.98 \AA, 1.21 \AA, 1.09 \AA, and 0.24 \AA ~for Ce, Ni, As, and O atoms, respectively. After obtaining the self-consistent electronic structure, maximally localized Wannier functions were constructed using the Wannier90 package \cite{Mostofi-Wannier90_CPC2008}. The anomalous Hall conductivity was subsequently calculated in the Wannier representation. A $120\times 120\times 120$ $k$-point mesh was adopted to evaluate the anomalous Hall conductivity of canted AFM and canted FM CeNiAsO.

\SMsection{SM \Rmnum{2}. S\lowercase{ample characterization}}

\newpage
\SMsection{SM \Rmnum{3}. R\lowercase{aw data of $^{75}$}A\lowercase{s} NQR \lowercase{and} NMR}

\begin{figure}[!ht]
\vspace*{-15pt}
\hspace*{0pt}
\includegraphics[width=17.5cm]{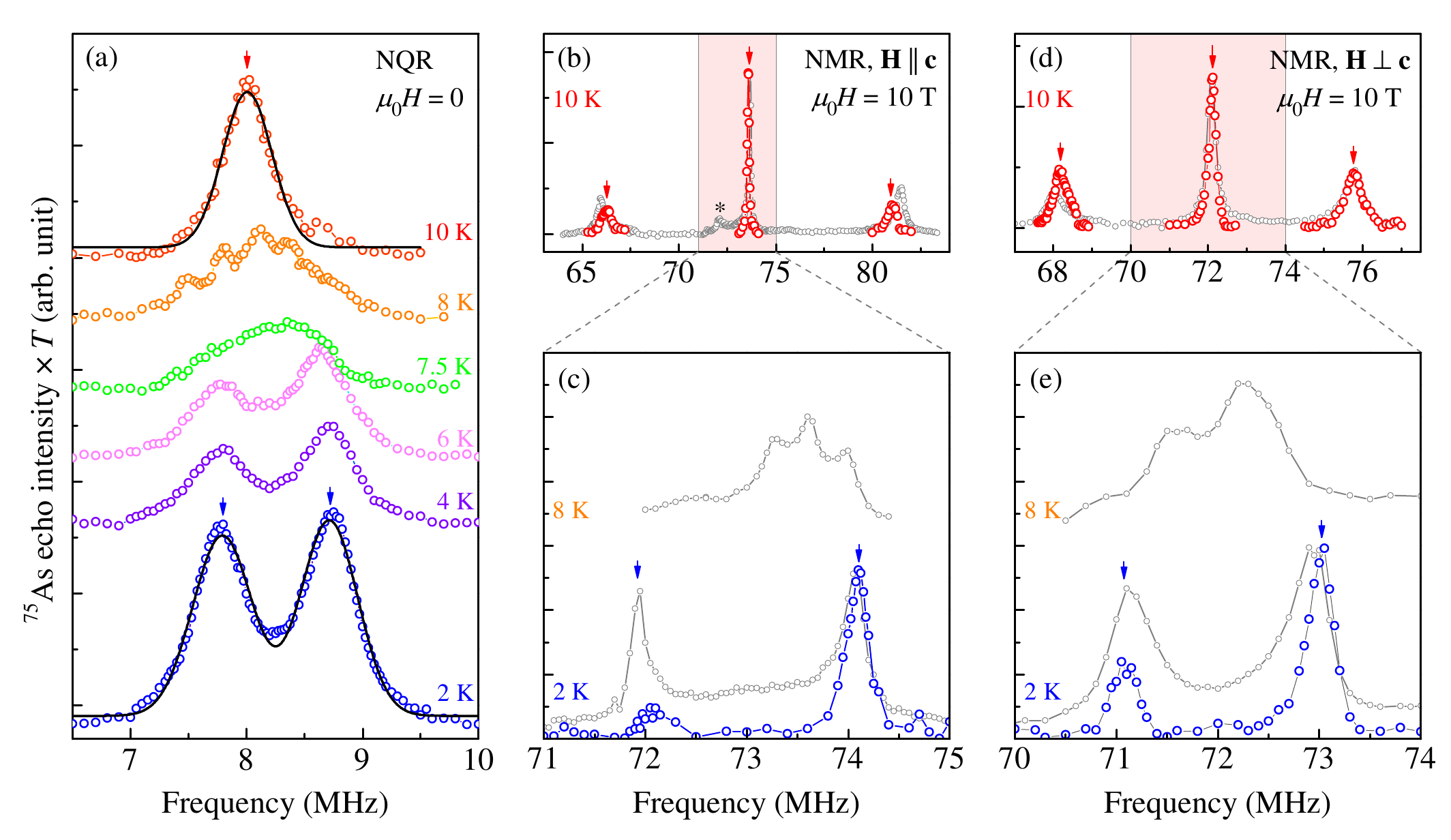}
\vspace*{-10pt}
\caption{(a) $^{75}$As NQR spectra of CeNiAsO for $T\leq10$ K. The black solid lines are the Gaussian fittings.  (b-c) $^{75}$As NMR spectra of CeNiAsO measured at external field $\mu_0H\approx10$ T, and $\mathbf{H}\parallel\mathbf{c}$. The gray circles are reproduced from our previous aligned-powder results \cite{LuFangjunCeNiAsO}, while the red and blue circles stand for the new results obtained from single crystals at 10 K and 2 K, respectively. The asterisk in panel (b) indicates a small amount of
nonaligned powders. (d-e), ibid, but for $\mathbf{H}\perp\mathbf{c}$. At 10 K in the paramagnetic phase, the resonance peaks are narrow; for $T_\text{N2}<T<T_\text{N1}$ in the incommensurate AFM phase, broad resonance peaks are observed due to the continuous distribution of internal field; for $T<T_\text{N2}$ in the commensurate AFM phase, well split resonance peaks are seen.}
\label{FigS2}
\end{figure}

\begin{figure}[!htp]
\vspace{-0pt}
\hspace{-0pt}
\includegraphics[width=16cm]{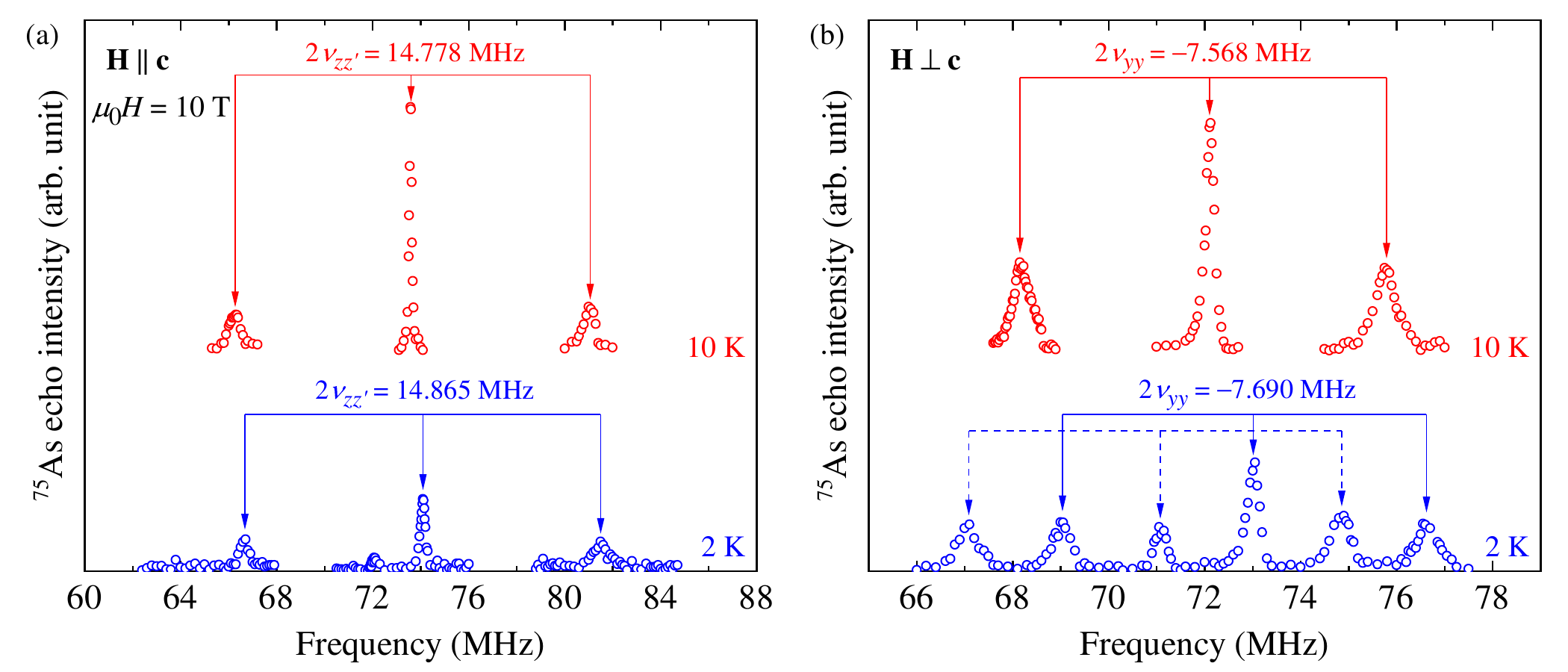}
\vspace{-10pt}
\caption{(a-b) $^{75}$As NMR spectra of single crystalline CeNiAsO for $\mathbf{H}\parallel\mathbf{c}$ and $\mathbf{H}\perp\mathbf{c}$, respectively. The external field $\mu_0 H=10$ T. Note that $\nu_{zz'}$ differs slightly from $\nu_{zz}$ (or $\nu_Q$) due to the small misalignment of $\mathbf{c}$ axis with respect to $\mathbf{H}$. }
\label{FigS3}
\end{figure}

\begin{figure}[!ht]
\vspace*{-0pt}
\hspace*{0pt}
\includegraphics[width=17cm]{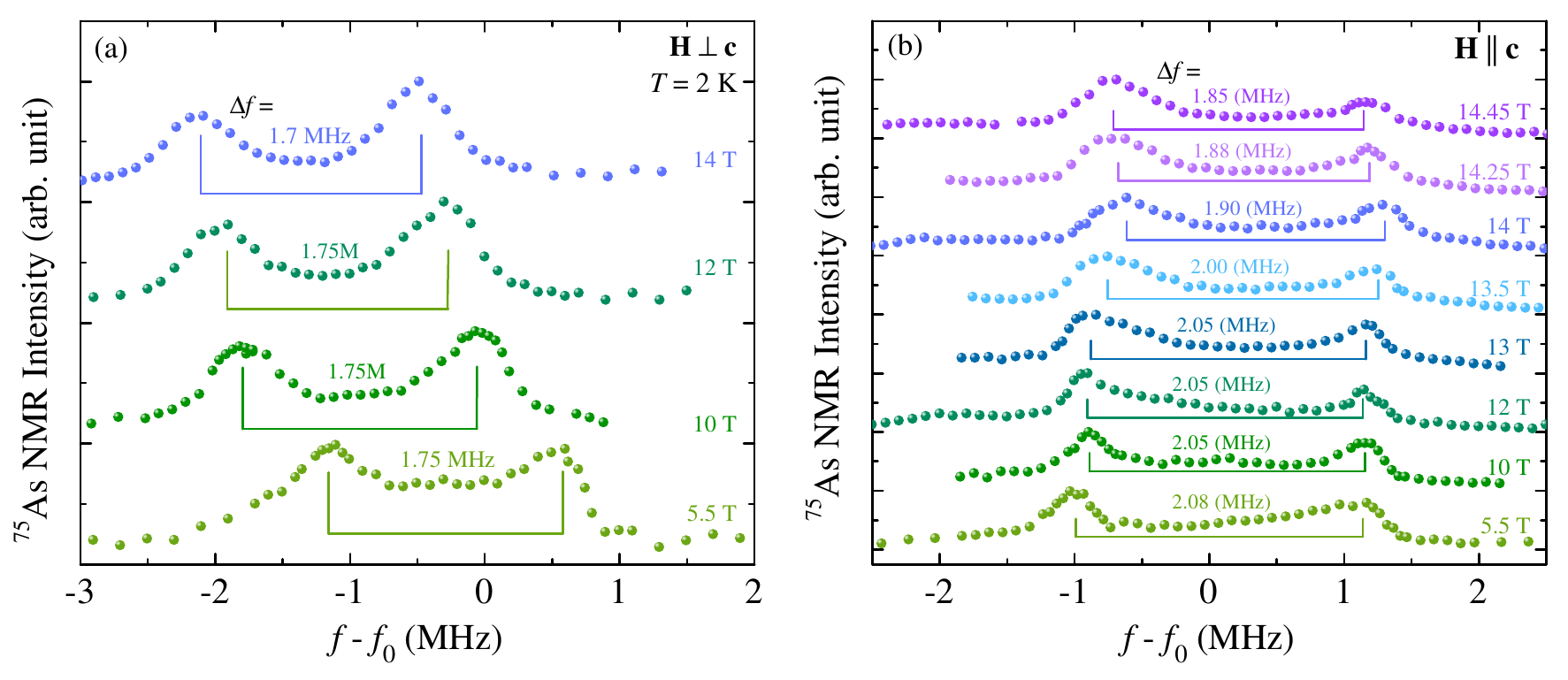}
\vspace*{-10pt}
\caption{NMR spectra of $^{75}$As central transition in aligned-powder CeNiAsO under different magnetic fields at 2 K. The reference frequency $f_0$ = $^{75}\gamma_n\mu_0H$. (a) $\mathbf{H}\perp\mathbf{c}$. (b) $\mathbf{H}\parallel\mathbf{c}$.
Below $\sim 13$ T, the peak splitting $\Delta f$ in both field directions remains essentially unchanged with field; above $\sim 13$ T, $\Delta f$ decreases substantially, suggesting that the magnetic moments start to be reorientated.}
\label{FigS4}
\end{figure}

\newpage
\SMsection{SM \Rmnum{4}.  D\lowercase{etermination of the internal field}}

To incorporate both NMR and NQR results, we write the total Hamiltonian as
\begin{equation}
\mathcal{H}_\text{tot}=\mathcal{H}_\text{Z}+\mathcal{H}_\text{int}+\mathcal{H}_\text{Q},
    \label{Htotal}
\end{equation}
where $\mathcal{H}_\text{Z}$ is the Zeeman term caused by the external field; $\mathcal{H}_\text{int}$ is the (Zeeman) term induced by the internal field ($\mathbf{B}_\text{int}$); and $\mathcal{H}_\text{Q}$ is the nuclear quadrupole interaction with the electric field gradient (EFG), taking the form
\begin{equation}
\mathcal{H}_Q=\frac{eQV_{zz}}{4I(2I-1)}[3\hat{I_z}^2-\mathbf{\hat{I}}^2+\eta(\hat{I_x}^2-\hat{I_y}^2)],
\label{HQ}
\end{equation}
in which $\mathbf{\hat{I}}=(\hat{I_x},\hat{I_y},\hat{I_z})$ is nuclear spin operator, $Q$ is nuclear quadrupole moment, and $\eta\equiv(V_{xx}-V_{yy})/V_{zz}$ is the asymmetry parameter with $V_{xx}$, $V_{yy}$ and $V_{zz}$ being the components of the EFG tensor.


According to the NMR experiments, we estimate that the amplitude of $\mathbf{B}_\text{int}$ is in the order of 0.1 T, in which case for the NQR measurements ($\mathcal{H}_\text{Z}=0$) the $\mathcal{H}_\text{int}$ can not be treated as a perturbation to $\mathcal{H}_\text{Q}$, or vise versa. A rigorous numerical calculation, therefore, is needed. Since the principal axes of these three terms are generally different, we diagonalize the total Hamiltonian via a unitary transformation
\begin{equation}
    \mathcal{H}=\mathbf{U}^\dagger~\mathcal{H}_{tot}~\mathbf{U},
    \label{EqS2}
\end{equation}
where $\mathbf{U}$ is the eigenvector matrix. The diagonal elements of the Hamiltonian represent the splitting of energy levels. Transitions only occur between specific energy levels, and their probabilities can be computed from
\begin{equation}
    \mathcal{P}=\mathbf{U}^\dagger~\hat{I}_{+}~\mathbf{U},
    \label{EqS3}
\end{equation}
where $\hat{I}_{+}=\hat{I}_x+i\hat{I}_y$. For $^{75}$As with $I=3/2$,
\begin{equation}
    \hat{I}_{+}=
    \left(
\begin{aligned}
    \begin{array}{cccc}
       0 & \sqrt{3} & 0 & 0\\
       0 & 0 & 2 & 0\\
       0 & 0 & 0 & \sqrt{3}\\
       0 & 0 & 0 & 0\\
    \end{array}
  \end{aligned}
\right).
    \label{EqS4}
\end{equation}

\begin{figure}[!htp]
\vspace{0pt}
\hspace{-0pt}
\includegraphics[width=11cm]{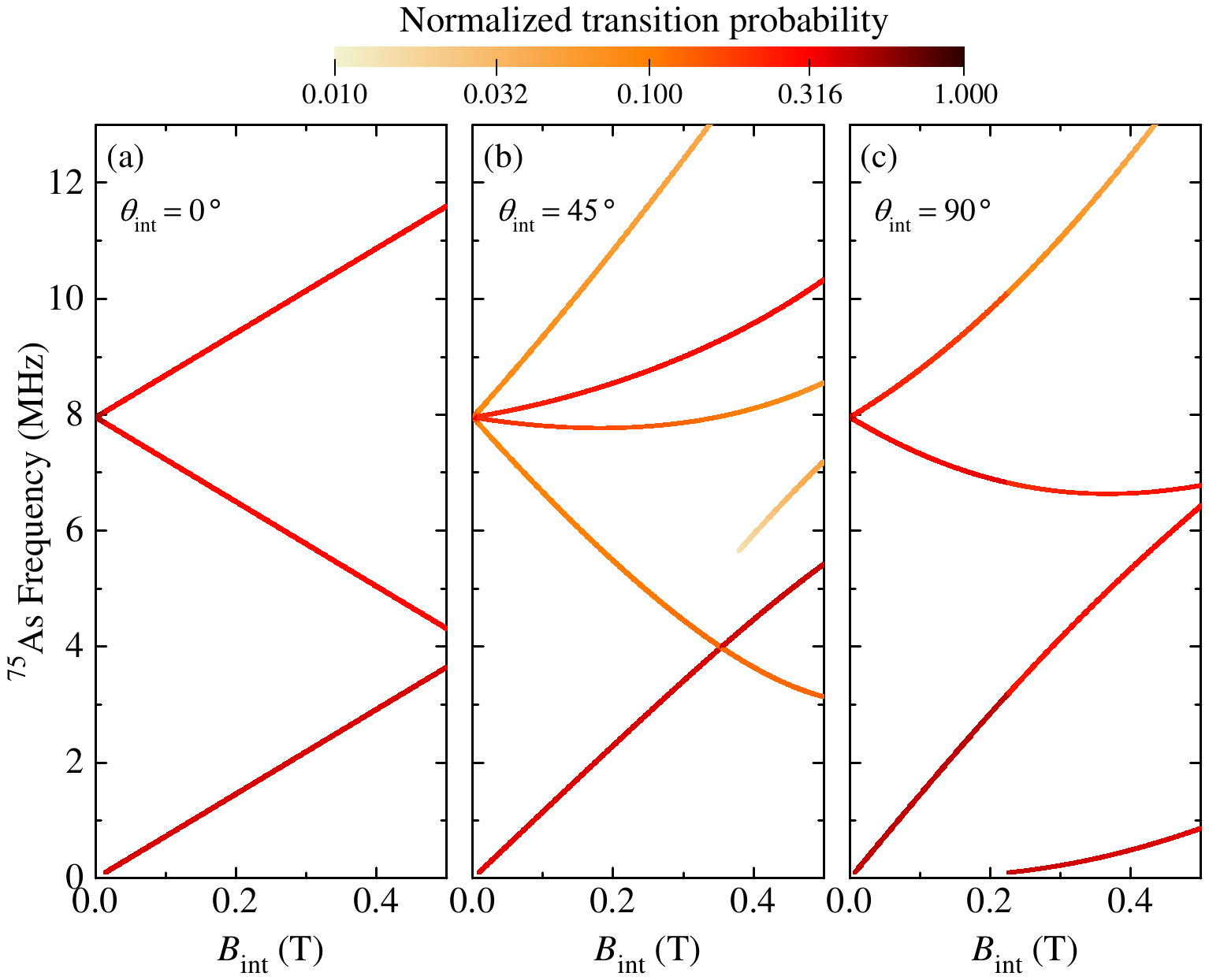}
\vspace{-10pt}
\caption{$^{75}$As NQR frequency as a function of $B_\text{int}$. The orientation of $\mathbf{B}_\text{int}$ is characterized by $\theta_\text{int}$ with respect to the principle axis of EFG.  (a) $\theta_\text{int} = 0$. (b) $\theta_\text{int} = 45~^{\circ}$. (c) $\theta_\text{int} = 90~^{\circ}$. The colormap represents the normalized transition probability.}
\label{FigS5}
\end{figure}

First, we can compute the $^{75}$As NQR frequency as a function of $B_\text{int}$ with specific $\mathbf{B}_\text{int}$ orientations $\theta_\text{int}$ with respect to the principle axis of EFG. In Fig.~\ref{FigS5}, we show the representative results for $\theta_\text{int}=0$, 45 and 90 $^\circ$, and the colormap indicates the normalized transition probabilities. For EFG principle axis parallel to the internal field ($\theta_\text{int} = 0~^{\circ}$), the energy level splitting and the resonance frequencies are linear with field. On the contrary, for EFG principle axis perpendicular to the internal field ($\theta_\text{int} = 90~^{\circ}$), these dependencies are nonlinear. As field increases, transitions between adjacent energy levels gradually become stronger, while transitions across energy levels weaken rapidly and finally disappear. For $\theta_\text{int} = 45~^{\circ}$, the situation is more complex: transitions between non-adjacent levels occur with significant probability, leading to multiple sub-branches of resonance frequency. 
The calculated results are then compared with the experiments, and we find that the best fitting parameters are $\theta_\text{int}\approx44~^\circ$ and $B_\text{int}=0.213$ T, seeing Fig.~2(a).

\begin{figure}[!htp]
\vspace{-0pt}
\hspace{-0pt}
\includegraphics[width=9cm]{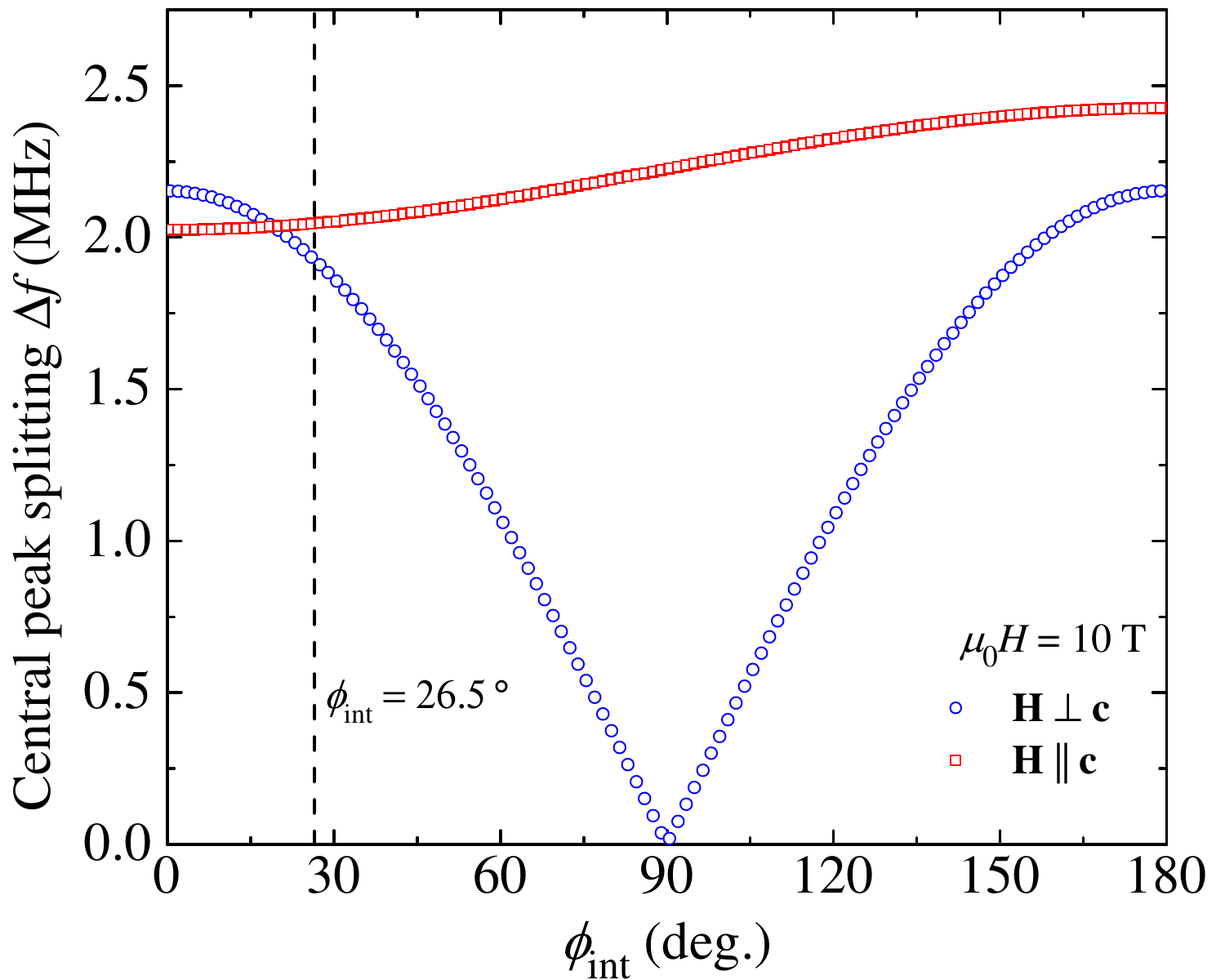}
\vspace{-5pt}
\caption{The calculated $^{75}$As NMR central peak splitting ($\Delta f$) as a function of the azimuthal angle of $\mathbf{B}_\text{int}$ with respect to $\mathbf{a}$ axis. Within the range of $[0, 180 ^\circ]$, $\Delta f$ for $\mathbf{H}\parallel\mathbf{c}$ increases monotonically with $\phi_\text{int}$, while the results for $\mathbf{H}\perp\mathbf{c}$ minimizes at $\phi_\text{int}=90 ^\circ$. It is found that only $\phi_\text{int} \approx 26.5 ^\circ$ can simultaneously fit both experimental results satisfactorily well.}
\label{FigS6}
\end{figure}

Next, the NMR frequencies (in the presence of external field $\mu_0H=10$ T) can also be simulated and compared with the experimental, seeing Fig.~2(b) and Fig.~\ref{FigS3}. At 10 K, given that the satellite-peak frequencies in single-crystal and polycrystalline NMR are essentially consistent for $\mathbf{H} \perp \mathbf{c}$, whereas the quadrupole splitting frequency measured in single-crystal NMR along $\mathbf{H} \parallel \mathbf{c}$ is relatively smaller than that in the polycrystalline case [cf Fig.~\ref{FigS2}(a)], we infer a slight misalignment of the external magnetic field away from the $\mathbf{c}$-axis. Fitting results indicate an angular deviation of approximately $5.5~^\circ$. Another salient feature is that, from 10 K to 2 K, the quadrupole splitting measured for $\mathbf{H}\perp\mathbf{c}$ (i.e. $\nu_{yy}$) increases much faster than that for $\mathbf{H}\parallel\mathbf{c}$ (i.e. $\nu_{zz'}$), seeing Fig.~\ref{FigS3}. This suggests the emergence of in-plane anisotropy due to magnetic ordering with $\mathbf{q}=(0.5,~0,~0)$, and yields a finite asymmetry parameter $\eta \approx 0.0102$. To better reconcile the NMR and NQR data, we attempted to fit all resonance peaks using a single $\nu_Q$ value; however, this turns out to be not very successful. Hence, we propose that under a strong magnetic field of 10 T, subtle modifications of the electric field gradient may occur, leading to distinct $\nu_Q$ values between NMR and NQR measurements. 
The simulated central peak splitting ($\Delta f$) for both $\mathbf{H}\parallel\mathbf{c}$ and $\mathbf{H}\perp\mathbf{c}$ are monotonic with the azimuthal angle of internal field $\phi_\text{int}$ in the range of $[0,~90^\circ]$, and this gives rise to $\phi_\text{int}=26.5~^{\circ}$ uniquely, seeing Fig.~\ref{FigS6}. 

\newpage
\SMsection{SM \Rmnum{5}. C\lowercase{alculations of hyperfine coupling tensors}}

\begin{figure}[!htp]
\vspace{-15pt}
\hspace{-0pt}
\includegraphics[width=10cm]{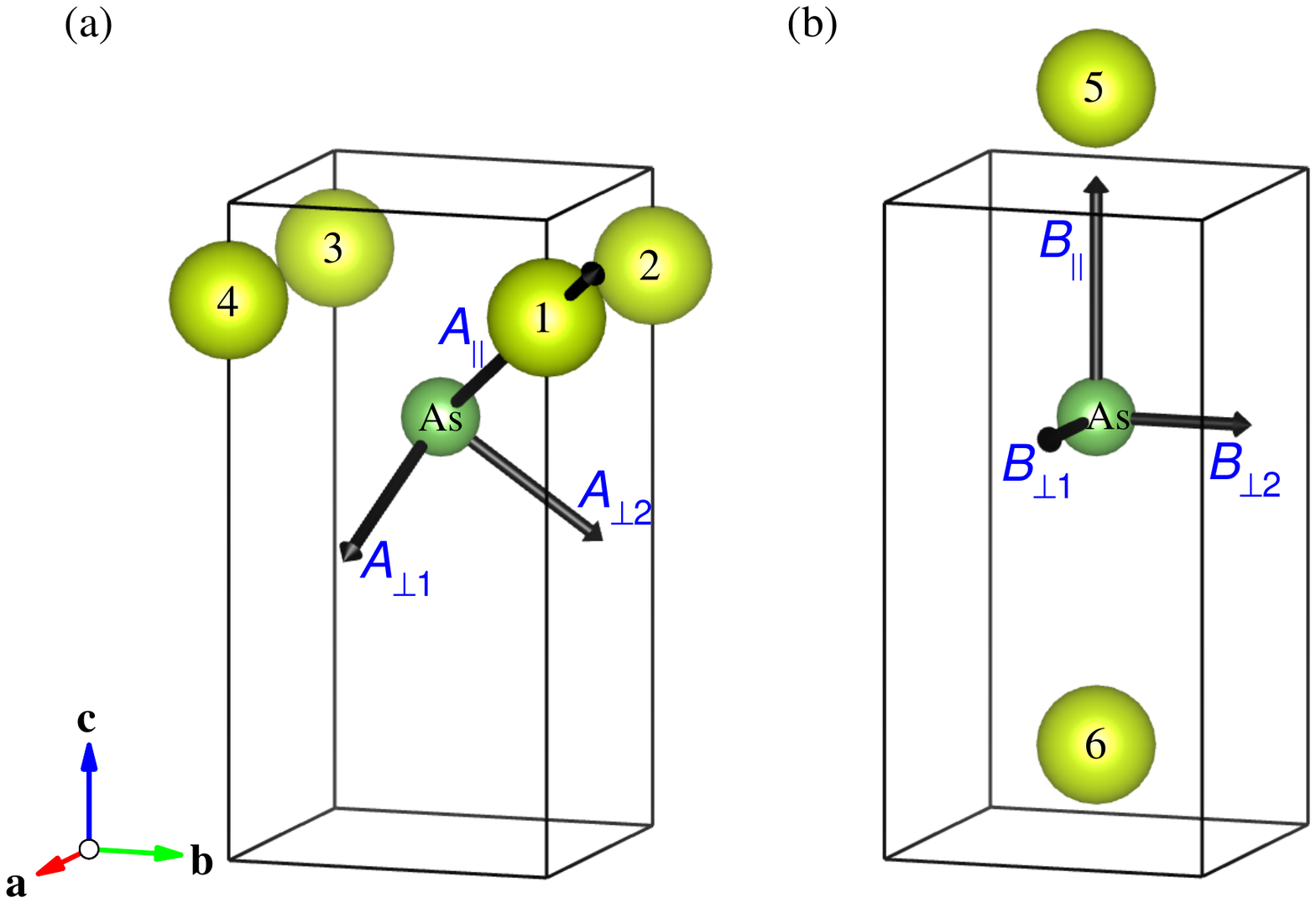}
\vspace{-5pt}
\caption{Local environment of As in CeNiAsO. Only Ce and As atoms are shown for clarity. There are six cerium neighbors surrounding the As, denoted as Ce1-6. The transferred hyperfine coupling relies on Ce-As orbital hybridization. (a) The principle axes of the transferred hyperfine tensors $\mathbb{A}_{i}$ of the As sites from the nearest-neighbor Ce (Take Ce1 as an example). (b) The principle axes of the transferred hyperfine tensors $\mathbb{A}_{i}$ of the As sites from the next-nearest neighbor Ce (as exemplified by Ce5). }
\label{FigS7}
\end{figure}

There are six cerium neighbors surrounding the As, as shown in Fig.~\ref{FigS7}. Among them, Ce1-4 are nearest neighbors, and Ce5-6 are next-nearest neighbors. In the lattice-coordinate system, the hyperfine coupling tensor of the nearest neighboring Ce1 atom writes as
\begin{equation}
\mathbb{A}_1 =
\left(
\begin{aligned}
    \begin{array}{ccc}
       A_{aa} & A_{ab} & A_{ac}\\
       A_{ba} & A_{bb} & A_{bc}\\
       A_{ca} & A_{cb} & A_{cc}\\
    \end{array}
  \end{aligned}
\right).
\label{Eq.S5}
\end{equation}
By symmetry, the tensors for the other three nearest-neighbor sites (Ce2-4) are
\begin{equation}
~\mathbb{A}_2 =
\left(
\begin{aligned}
    \begin{array}{ccc}
       A_{aa} & -A_{ab} & -A_{ac}\\
       -A_{ba} & A_{bb} & A_{bc}\\
       -A_{ca} & A_{cb} & A_{cc}\\
    \end{array}
  \end{aligned}
\right),
~\mathbb{A}_3 =
\left(
\begin{aligned}
    \begin{array}{ccc}
        A_{aa} & A_{ab} & -A_{ac}\\
       A_{ba} & A_{bb} & -A_{bc}\\
       -A_{ca} & -A_{cb} & A_{cc}\\
    \end{array}
  \end{aligned}
\right),
~\mathbb{A}_4 =
\left(
\begin{aligned}
    \begin{array}{ccc}
       A_{aa} & -A_{ab} & A_{ac}\\
       -A_{ba} & A_{bb} & -A_{bc}\\
       A_{ca} & -A_{cb} & A_{cc}\\
    \end{array}
  \end{aligned}
\right).
\label{Eq.S6}
\end{equation}

For the next-nearest neighbors (Ce5-6), the hyperfine coupling tensors are
\begin{equation}
\mathbb{A}_5 =
\left(
\begin{aligned}
    \begin{array}{ccc}
       B_{aa} & B_{ab} & B_{ac}\\
       B_{ba} & B_{bb} & B_{bc}\\
       B_{ca} & B_{cb} & B_{cc}\\
    \end{array}
  \end{aligned}
\right),
~\mathbb{A}_6 =
\left(
\begin{aligned}
    \begin{array}{ccc}
       B_{aa} & B_{ab} & -B_{ac}\\
       B_{ba} & B_{bb} & -B_{bc}\\
       -B_{ca} & -B_{cb} & B_{cc}\\
    \end{array}
  \end{aligned}
\right).
\label{Eq.S7}
\end{equation}

According to the previous neutron-scattering results \cite{ShanWu2019}, in the commensurate AFM phase with $\mathbf{q}=(0.5,0,0)$, the moments $\mathbf{m}_1=-\mathbf{m}_2=-\mathbf{m}_3=\mathbf{m}_4$; $m_{5x}=m_{1x}$, and $m_{5y}=-m_{1y}$. Putting these together, the internal field at As site can be obtained,
\begin{equation}
\mathbf{B}_\text{int}=(4A_{ac}m_z+2B_{aa}m_x-2B_{ab}m_y,~2B_{ba}m_x-2B_{bb}m_y,~4A_{ca}m_x\pm2B_{cc}m_z),
\label{Eq.B_int}
\end{equation}
where ``$+/-$" denotes that the upper and lower Ce layers in a single CeO slab have the same/opposite $m_z$ components, viz $m_{1z}=\pm m_{5z}$.

In the bond coordinate (cf Fig.~\ref{FigS7}), the tensors $\tilde{\mathbb{A}}_i$ are diagonal, e.g.
\begin{equation}
\tilde{\mathbb{A}}_1 =
\left(
\begin{aligned}
    \begin{array}{ccc}
       A_{\perp 1} & 0 & 0\\
       0 & A_{\perp 2} & 0\\
       0 & 0 & A_{\parallel}\\
    \end{array}
  \end{aligned}
\right),
~\tilde{\mathbb{A}}_5 =
\left(
\begin{aligned}
    \begin{array}{ccc}
       B_{\perp 1} & 0 & 0\\
       0 & B_{\perp 2} & 0\\
       0 & 0 & B_{\parallel}\\
    \end{array}
  \end{aligned}
\right),
\label{Eq.S8}
\end{equation}
where the elements $A_\parallel$ and $B_\parallel$ mean the hyperfine coupling along the Ce-As bonding, while $A_{\perp1}$, $A_{\perp2}$, $B_{\perp1}$ and $B_{\perp2}$ are the components for perpendicular to the bonds. $\mathbb{A}_i$ and $\tilde{\mathbb{A}}_i$ are connected and converted through coordinate transformation,
 \begin{equation}
\mathbb{A}_i = \mathbf{R}^{-1}~\tilde{\mathbb{A}}_i~\mathbf{R},
\label{Eq.S9}
\end{equation}
where $\mathbf{R}$ is the coordinate transformation matrix. In particular, for Ce5 and Ce6, the two coordinates are equivalent, therefore,
\begin{equation}
\mathbb{A}_5 = \tilde{\mathbb{A}}_5=
\left(
\begin{aligned}
    \begin{array}{ccc}
       B_{\perp 1} & 0 & 0\\
       0 & B_{\perp 2} & 0\\
       0 & 0 & B_{\parallel}\\
    \end{array}
  \end{aligned}
\right).
\label{Eq.S10_B}
\end{equation}
For $\mathbb{A}_1$, it can be obtained via rotating $\tilde{\mathbb{A}}_1$ by approximately $-60^{\circ}$ about the ($-1~1~0$) direction, i.e.,
\begin{equation}
\mathbb{A}_1 =
\left(
\begin{aligned}
    \begin{array}{ccc}
       \frac{9A_{\perp 1}}{16}+\frac{A_{\perp 2}}{16}+\frac{3 A_{\parallel}}{8} & \frac{3A_{\perp 1}}{16}+\frac{3A_{\perp 2}}{16}-\frac{3 A_{\parallel}}{8} & \frac{3\sqrt{6}A_{\perp 1}}{16}-\frac{\sqrt{6}A_{\perp 2}}{16}-\frac{\sqrt{6}A_{\parallel}}{8}\\
       \frac{3A_{\perp 1}}{16}+\frac{3A_{\perp 2}}{16}-\frac{3 A_{\parallel}}{8} & \frac{A_{\perp 1}}{16}+\frac{9A_{\perp 2}}{16}+\frac{3 A_{\parallel}}{8} & \frac{\sqrt{6}A_{\perp 1}}{16}-\frac{3\sqrt{6}A_{\perp 2}}{16}+\frac{\sqrt{6}A_{\parallel}}{8}\\
       \frac{3\sqrt{6}A_{\perp 1}}{16}-\frac{\sqrt{6}A_{\perp 2}}{16}-\frac{\sqrt{6}A_{\parallel}}{8} & \frac{\sqrt{6}A_{\perp 1}}{16}-\frac{3\sqrt{6}A_{\perp 2}}{16}+\frac{\sqrt{6}A_{\parallel}}{8} & \frac{3A_{\perp 1}}{8}+\frac{3A_{\perp 2}}{8}+\frac{A_{\parallel}}{4}\\
    \end{array}
  \end{aligned}
\right).
\label{Eq.S10_A}
\end{equation}
Considering the tetragonal symmetry, $A_{aa}=A_{bb}$, $B_{aa}=B_{bb}$. Therefore, $A_{\perp 1}= A_{\perp 2}=A_{\perp}$, and $B_{\perp 1}= B_{\perp 2}=B_{\perp}$. Eq.~(\ref{Eq.S10_B}-\ref{Eq.S10_A}) can thus be further simplified.

\newpage
\SMsection{SM \Rmnum{6}. R\lowercase{efinement of the magnetic structure}}

In order to obtain the possible magnetic structure, we need to find out the orientations of each Ce moment - a configuration that can simultaneously satisfy all the existing NMR and NQR results. Here we have to make a few appropriate assumptions. First of all, we assume that the magnetic structure (and also the $\mathbf{B}_\text{int}$) of CeNiAsO is robust to external field up to 10 T, which seems to reasonable because the field dependent magnetization curve $M(H)$ remains linear and is far away from saturation even for field as high as 58 T, seeing Fig.~\ref{FigS1}(c). Additional evidence for this can be provided by the field dependent NMR shown in Fig.~\ref{FigS4}, from which we notice that the peak splitting $\Delta f$ does not change much below $\sim 13$ T.

The Knight shift $K$ along the principal axis can be expressed in terms of $\mathbb{A}_i$ and the magnetic susceptibility $\chi(T)$,
\begin{equation}
K_\alpha(T)\approx(4A_{\alpha\alpha}+2B_{\alpha\alpha})\chi(T),
\label{Eq.KChi}
\end{equation}
where $\alpha=a,b,c$. Eq.~(\ref{Eq.KChi}) omitted  on-site hyperfine interaction to the conduction electron spin, whose effect is much weaker than the localized $4f$ electrons.  From the previous NMR experiments, the parameters $A_{\text{hf},c}^\text{exp}=0.86(2)$ T/$\mu_\text{B}$ and $A_{\text{hf},ab}^\text{exp}=-0.42(9)$ T/$\mu_\text{B}$ were obtained through Clogston-Jaccarino plot $K$ vs $\chi$ \cite{LuFangjunCeNiAsO}. Neutron scattering gives the magnitude of ordered moment $m=0.37(5)~\mu_\text{B}$ \cite{ShanWu2019}, and
\begin{equation}
m_x^2+m_y^2+m_z^2=m^2.
\label{Eq.S12}
\end{equation}
According to the NQR analysis shown in Fig.~2(a), we already get $B_\text{int}=0.213$ T, polar angle $\theta_\text{int}=44~^{\circ}$, and azimuthal angle $\phi_\text{int}=26.5~^{\circ}$. Therefore,
\begin{equation}
B_{\text{int},x}^2+B_{\text{int},y}^2+B_{\text{int},z}^2=B_\text{int}^2,
\label{Eq.S13}
\end{equation}
\begin{equation}
B_{\text{int},x}^2+B_{\text{int},y}^2=B_{\text{int},z}^2\tan^2{\theta_\text{int}},
\label{Eq.S14}
\end{equation}
and
\begin{equation}
B_{\text{int},y}^2/B_{\text{int},x}^2=\tan^2{\phi_\text{int}}.
\label{Eq.S15}
\end{equation}



By solving Eq.~(\ref{Eq.B_int}) and (\ref{Eq.S10_A}-\ref{Eq.S15}) simultaneously, we can obtain possible hyperfine coupling tensors for different magnetic moment directions characterized by $\theta_\mathbf{m}$ and $\phi_\mathbf{m}$. The out-of-plane hyperfine coupling constant $A_{\text{hf},c}$ is imposed as a constraint, whereas the in-plane constant $A_{\text{hf},ab}$ serves as a validation criterion. The theoretical value $A_{\text{hf},ab}^\text{cal}$ can be calculated from the obtained hyperfine coupling tensor via Eq.~(\ref{Eq.KChi}) and then compared with the experimental $A_{\text{hf},ab}^\text{exp}$. The goodness of the fitting is evaluated by $\log_{10}R_p$, where
\begin{equation}
\begin{aligned}
R_p(\theta_\mathbf{m},~\phi_\mathbf{m})&={(A_{\text{hf},ab}^\text{cal}-A_{\text{hf},ab}^\text{exp})^2}\\
&={[4A_{aa}(\theta_\mathbf{m},~\phi_\mathbf{m})+2B_{aa}(\theta_\mathbf{m},~\phi_\mathbf{m})-A_{\text{hf},ab}^\text{exp}]^2}.
\end{aligned}
\label{Eq.S19}
\end{equation}
The optimal parameters $(\theta_\mathbf{m},~\phi_\mathbf{m})$ should lead to the minimal $\log_{10}R_p$, as is shown in Fig.~2(c).

\newpage
\SMsection{SM \Rmnum{7}. R\lowercase{aw data of} H\lowercase{all effect}}

\begin{figure}[!htp]
\vspace{-15pt}
\hspace{-7pt}
\includegraphics[width=18.2cm]{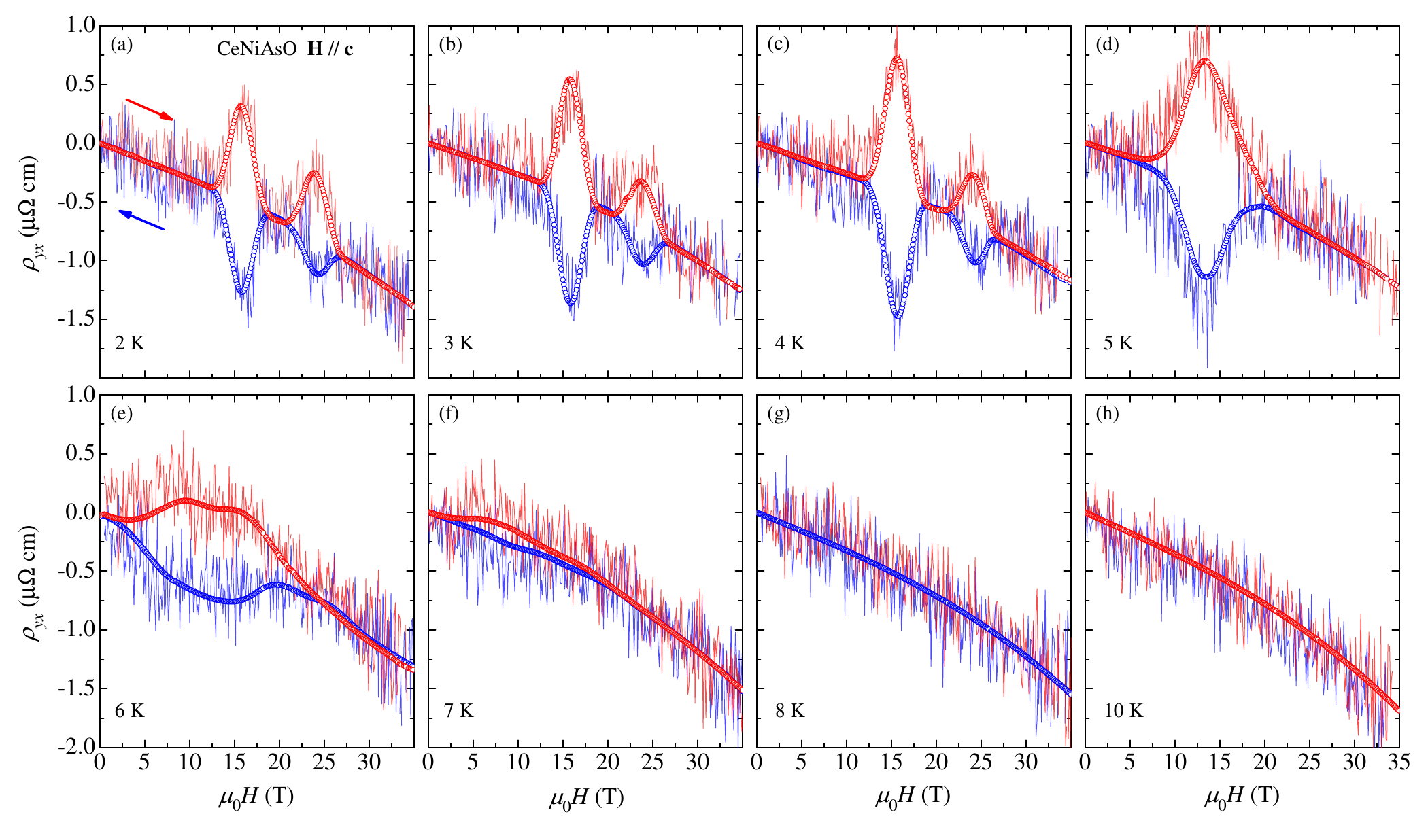}
\vspace{-25pt}
\caption{Field dependence of Hall resistivity $\rho_{yx}$ in CeNiAsO single crystal at selected low temperatures. The open symbols are smoothed curves, serving as guides to the eye. (a) $T = 2$ K. (b) $T = 3$ K. (c) $T = 4$ K. (d) $T = 5$ K. (e) $T = 6$ K. (f) $T = 7$ K. (g) $T = 8$ K. (h) $T = 10$ K.}
\label{FigS8}
\end{figure}

In Fig.~\ref{FigS8} we present the raw data of Hall effect of CeNiAsO single crystal for $\mathbf{H}\parallel\mathbf{c}$. Due to the good metallicity of CeNiAsO, the Hall signal is very small. The raw data were smoothed and fitted with great care. At low temperatures well below $T_\text{N2}$, two distinct hysteresis loops are clearly observed in the Hall resistivity $\rho_{yx}$ upon sweeping the magnetic field upward and downward [Fig.~\ref{FigS8}(a-c)]. When approaching $T_\text{N2}$, these anomalies are strongly suppressed, evolving into a weak single-loop response at 5-6 K [Fig.~\ref{FigS8}(d-e)]. Such anomalous Hall effect (AHE) dies away for temperature above $T_\text{N2}$ [Fig.~\ref{FigS8}(f)], in both of the incommensurate AFM [Fig.~\ref{FigS8}(g)] and paramagnetic [Fig.~\ref{FigS8}(h)] regimes. The systematic temperature evolution indicates that the anomalous Hall response is closely associated with the commensurate AFM phase that is deemed as a $p$-wave magnetic state \cite{P_wave_magnets_2023, Chakraborty2025_NC, ZhaoPRB2025, LuoX-OddParityMag}.

To extract the AHE component, we express the total Hall resistivity as follows,
\begin{equation}
\rho_{yx}(H)=\rho_{yx}^\text{O}+\rho_{yx}^\text{A},
\end{equation}
where $\rho_{yx}^\text{O}$ and $\rho_{yx}^\text{A}$ are the ordinary and anomalous Hall resistivities, respectively. In the high-field limit where the condition $\omega_\text{c}\tau_\text{tr} \ll 1$ no-longer holds ($\omega_c$ is cyclotron frequency; $\tau_\text{tr}$ is scattering time), $\rho_{yx}^\text{O}(H)$ becomes superlinear and can be fitted to a superposition of linear ($H$) and quadratic ($H^2$) terms \cite{Nair-CeIrIn52008, HeX-Ce3TiSb5_Hall2025}. After subtracting the background contribution of $\rho_{yx}^\text{O}$, the anomalous contribution $\rho_{yx}^\text{A}$ can be obtained, as shown in Fig.~4(b).

\newpage
\SMsection{SM \Rmnum{8}. T\lowercase{he direction of the spin polarization axis}}

\begin{figure}[!htp]
\vspace{-10pt}
\hspace{-0pt}
\includegraphics[width=8.5cm]{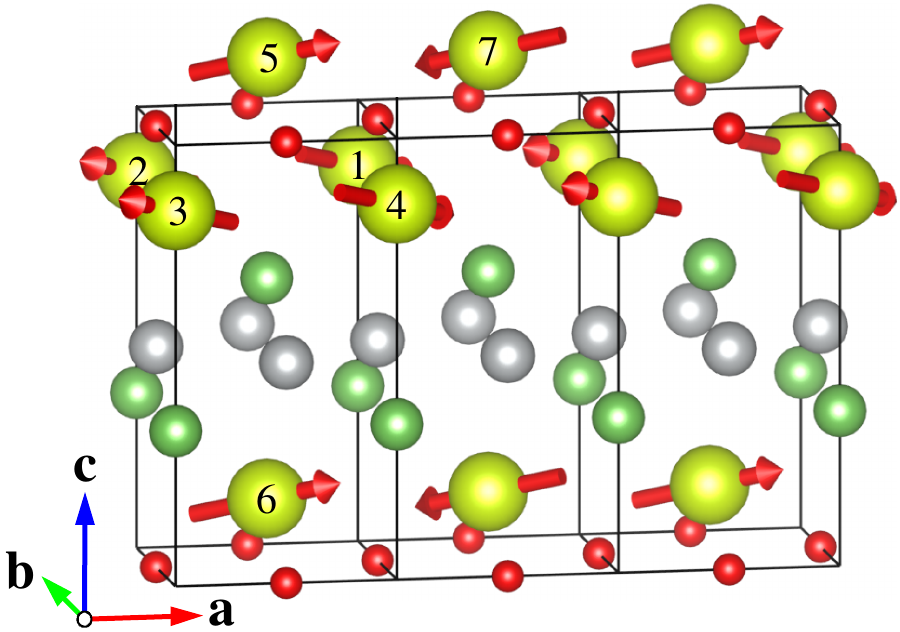}
\vspace{-0pt}
\caption{Tilted coplanar CAFM$_z$ order of CeNiAsO. }
\label{FigS9}
\end{figure}

To determine whether the magnetic moments form a non-coplanar spin configuration, we analyze the geometrical relation among the 4 kinds of magnetic moments in the CAFM$_z$ order of CeNiAsO:
\begin{equation}
	\begin{aligned}
		\mathbf{m}_1&=(m_x,~m_y,~-m_z),\\
		\mathbf{m}_2&=(-m_x,~-m_y,~m_z),\\
		\mathbf{m}_5&=(m_x,~-m_y,~m_z),\\
		\mathbf{m}_7&=(-m_x,~m_y,~-m_z),
	\end{aligned}
\end{equation}
with $m_x=|\boldsymbol{m}|\cos\theta_{\bm m}\cos\phi_{\bm m}$, $m_y=|\boldsymbol{m}|\cos\theta_{\bm m}\sin\phi_{\bm m}$, and $m_z=|\boldsymbol{m}|\sin\theta_{\bm m}$. Note that $\theta_\mathbf{m}$ is with respect to the $\mathbf{ab}$ plane. The normal vector ($\mathbf{n}$) of the plane spanned by $\mathbf{m}_1$, $\mathbf{m}_2$, and $\mathbf{m}_5$ can be obtained from their cross product:
\begin{equation}
	\mathbf{n}=\mathbf{m}_2\times\mathbf{m}_5
	=(0,~-2m_xm_z,~-2m_xm_y).
\end{equation}
The remaining magnetic moment satisfies
\begin{equation}	\mathbf{n}\cdot\mathbf{m}_7=(0,~-2m_xm_z,~-2m_xm_y)\cdot(-m_x,~m_y,~-m_z)=0,
\end{equation}
demonstrating that all four magnetic moments lie within the same plane. Therefore, this magnetic configuration is coplanar. The normal direction of the magnetic plane is given by
\begin{equation}
	\hat{\mathbf{n}}=\frac{\mathbf{n}}{|\mathbf{n}|}=\frac{(0,~m_z,~m_y)}{\sqrt{m_z^2+m_y^2}}\approx (0,~0.231,~0.973).
\end{equation}
This magnetic canting establishes a new spin plane, whose normal direction is rotated from the crystallographic $\mathbf{c}$ axis to $\hat{\mathbf{n}}$.

\end{document}